\documentclass[final,3p,times,authoryear]{elsarticle}
\usepackage{graphicx}
\usepackage{amssymb}
\usepackage{newtxtext}
\usepackage{newtxmath}
\usepackage{natbib}
\usepackage{xcolor}
\usepackage{hyperref}
\usepackage{cleveref} 
\usepackage{float}

\usepackage{amsmath}
\usepackage{booktabs}
\usepackage{tabularx}
\usepackage{array} 
\usepackage{caption}  
\usepackage{algorithm} 
\usepackage{algpseudocode}

\hypersetup{
    colorlinks=true,
    citecolor=blue,    
    linkcolor=blue,    
    filecolor=blue,    
    urlcolor=blue      
}

\crefname{equation}{equation}{Eqs.}
\Crefname{equation}{equation}{Eqs.}
\crefname{figure}{figure}{figures}
\Crefname{Figure}{Figure}{Figures}
\crefname{algorithm}{Algorithm}{Algorithms}
\Crefname{algorithm}{Algorithm}{Algorithms}
\crefname{section}{section}{sections}
\Crefname{Section}{Section}{Sections}
\crefname{table}{table}{tables}
\Crefname{Table}{Table}{Tables} 

\begin{document}

\begin{frontmatter}

\title{Strategies for energy-efficient flow control leveraging deep reinforcement learning} 

\author{Wang Jia, Hang Xu} 

\affiliation{organization={State Key Laboratory of Ocean Engineering, School of Ocean and Civil Engineering, Shanghai Jiao Tong University},
            city={Shanghai},
            postcode={200240}, 
            country={China}}

\begin{abstract}
This study investigates active flow control in two-dimensional flows at a Reynolds number of 100 using Deep Reinforcement Learning (DRL). We utilize DRL to develop flow control strategies that enhance energy efficiency and minimize energy consumption, thereby addressing the limitations of traditional methods.
We find that the optimal jet placement for both square and circular cylinders is at the main flow separation point, achieving the best balance between energy efficiency and control effectiveness.
For the circular cylinder, positioning the jet at approximately 105° from the stagnation point requires only 1\% of the inlet flow rate and achieves an 8\% reduction in drag, with energy consumption one-third of that at other positions. For the square cylinder, placing the jet near the rear corner requires only 2\% of the inlet flow rate, achieving a maximum drag reduction of 14.4\%, whereas energy consumption near the front corner is 27 times higher, resulting in only 12\% drag reduction.
In multi-action control, the convergence speed and stability are lower compared to single-action control, but activating multiple jets significantly reduces initial energy consumption and improves energy efficiency. 
Physically, the interaction of the synthetic jet with the flow generates new vortices that modify the local flow structure, significantly enhancing the cylinder's aerodynamic performance.
Our control strategy achieves a superior balance between energy efficiency and control performance compared to previous studies, underscoring its significant potential to advance sustainable and effective flow control.
\end{abstract}







\end{frontmatter}

\section{INTRODUCTION}

Active flow control (AFC) involves regulating and controlling fluid flow through external mechanisms such as injection, suction, or vibration to optimize flow characteristics and enhance system efficiency, performance, and stability. Despite significant advancements in this field over the past few decades, there remain unresolved challenges, particularly in addressing the nonlinear complexities inherent in fluid dynamics \cite{collis2004issues}. Fluid dynamics problems are intrinsically nonlinear, especially when dealing with turbulent and separated flows, where the flow behavior is extremely complex and difficult to control accurately using traditional analytical or numerical methods \cite{cattafesta2011actuators}. These challenges are intensified in high-dimensional, multi-scale problems, making global optimization, real-time control, robustness, and energy efficiency difficult to achieve \cite{BEWLEY200121}.

The rapid development of artificial intelligence has brought new vitality to the field of flow control. Reinforcement learning (RL) has emerged as a key machine learning algorithm that effectively addresses complex decision-making and control problems \cite{annurevfluid,ClosedLoop}. RL agents continuously adjust their behavior strategies in response to environmental feedback, making RL a powerful tool for autonomous learning and decision-making \cite{arulkumaran2017deep}. Deep learning (DL) has strong expressive capabilities, allowing it to approximate complex nonlinear functions \cite{Janiesch2021}. When combined with RL to form deep reinforcement learning (DRL), this approach leverages DL's feature extraction capabilities and RL's strategy optimization strengths \cite{franccois2018introduction}. 
DRL originated in the mid-20th century, with deep learning and computational power driving its development. The 2013 introduction of Deep Q-Network by Google DeepMind was a major breakthrough, catalyzing rapid growth \cite{mnih2013playing}. Innovations like PPO  have further enhanced DRL's stability and efficiency, enabling its application in fields such as autonomous navigation and manufacturing \cite{li2018,schulman2017}.

Given the exceptional decision-making capabilities of reinforcement learning, its methods and derivative technologies are seeing widespread development in fluid mechanics, covering various areas such as turbulence modeling, flow prediction, shape optimization, and flow control \cite{rabault2020deep,Viquerat,Vignon2023}. 
In the domain of turbulence modeling, \cite{novati2021automating} leveraged multi-agent reinforcement learning to automate the discovery of turbulence models, while \cite{KURZ2023109094} utilized a multi-agent reinforcement learning framework to identify the optimal eddy viscosity model for implicitly filtered large eddy simulations. 
In the field of shape optimization, \cite{VIQUERAT2021110080} spearheaded the application of DRL for direct shape optimization, demonstrating that artificial neural networks (ANN) could autonomously generate optimal shapes within a constrained time frame and without prior knowledge, given an appropriate reward mechanism.
In the context of flow control, \citeauthor{rabault2019artificial} introduced the first application of an ANN trained via DRL for flow control, successfully reducing drag by 8\%.

AFC techniques combining computational fluid dynamics (CFD) with DRL have been extensively studied for robustness across Reynolds numbers ($Re$), flow separation control, parallelization, algorithm comparisons, and probe optimization.
Firstly, DRL controllers have shown strong generalization across various $Re$. 
\citeauthor{tangRobust2020} achieved significant drag and lift reductions in cylinder flows, while \citeauthor{jia2024robust} demonstrated the DRL agent's ability to fully suppress vortex shedding across multiple $Re$.
\citeauthor{renApplying} and \citeauthor{wang2024dynamic} further explored DRL in higher $Re$ and 3D flows.
Secondly, DRL algorithms have been successfully applied to control complex flow separation phenomena across various configurations, including fluidic pinball systems \cite{Feng2023}, square cylinders \cite{jia2024robust}, airfoils \cite{wangairfoil}, elliptical cylinders \cite{wang2024deep}, and flat plates \cite{wang2024deep}, showcasing their versatility and robustness.  
Thirdly, significant advancements have been achieved in refining DRL algorithms and optimizing probe placements, as highlighted by \citeauthor{liReinforcementlearning,Xia2024,suarez2024,paris2021robust}
Additionally, parallelization has proven critical for enhancing the efficiency of DRL-based flow control, with studies such as \citeauthor{rabaultAccelerating}, \citeauthor{jia2024optimal} and \citeauthor{wangDRLinFluids} demonstrating its potential to accelerate training and strategy discovery.

Despite prior studies exploring DRL-based AFC from various perspectives, research in the field of synthetic jet technology has identified that the placement of jet actuators plays a crucial role in determining control effectiveness and energy consumption \cite{ZHU2019468,act7040077}.
The strategic layout of synthetic jets plays a pivotal role in determining their interaction with boundary layer flows, thereby influencing both control performance and energy efficiency \cite{LIN2002389,QAYOUM20105035,202410}. 
As a highly effective flow control method, the optimal positioning of synthetic jets directly dictates their interaction with the boundary layer and the main flow, affecting aerodynamic properties such as drag reduction and lift modulation \cite{CHEN2022111840,202410,GIRFOGLIO2015512}. 
Moreover, different jet locations lead to varying levels of energy utilization efficiency, where improper placement may result in increased energy consumption with limited control benefits \cite{SAIDUR20101135,ZUCKERMAN2006565}. 
Investigating the coupling between synthetic jet placement and control performance is essential for understanding mechanisms, yet systematic evaluation of this balance remains limited.

Recent studies have begun to explore the influence of synthetic jet actuator placement on flow control. 
\citeauthor{chen2023deep} demonstrated that placing actuators near the rear corners of a square cylinder at $Re = 100$ improved vortex shedding suppression.
At higher $Re$, \citeauthor{yan2023stabilizing} showed that positioning actuators near the front corners enhanced performance at $Re = 2000$.
However, two key issues remain: firstly, these studies primarily focus on control effectiveness without providing a detailed analysis of the underlying physical mechanisms. 
Secondly, their qualitative findings on optimal jet placement vary across different $Re$, leading to inconsistent results \cite{chen2023deep,yan2023stabilizing}. 
This has motivated research exploring the impact of synthetic jet location and width on flow control performance at $ Re = 100 $ and 500, utilizing an entropy-regularized reinforcement learning approach \cite{jia2024jetsactuator}.
Despite these advances, there remains a lack of comprehensive research on the sensitivity of synthetic jet placement to flow separation suppression in bluff body, as well as fluid mechanics-based qualitative  coupled with quantitative analysis.

We note that \citeauthor{yan2024aero} highlighted the importance of multiple jet actuators, where configurations with multiple independent actuators demonstrated optimal performance. Thus, we have developed a strong interest in comparing the control performance of multiple synthetic jets operating simultaneously with that of single jets working independently.
Firstly, the synergy among multiple jets can induce nonlinear flow behaviors, resulting in control effects that differ significantly from those achieved by a single actuator. Evaluating these performance differences can reveal the potential and limitations of synthetic jets in flow control. Secondly, as the number of actuators increases, the interactions between jets and the complexity of energy distribution also rise, potentially impacting energy consumption and control efficiency. 
A systematic comparison of multiple versus single-pair synthetic jet activation offers insights into achieving efficient, energy-saving flow control.

Given the identified challenges and research gaps in multiple aspects, we conduct a comprehensive analysis of synthetic jet placement to evaluate its impact on flow control performance and energy efficiency.
We have designed five synthetic jet placements for cylindrical bodies, resulting in a total of nine jet configuration schemes, including five control strategies for single synthetic jet actions and four for simultaneous multi-jet actions. Additionally, for square cylinders, we designed four jet placements, leading to seven jet configuration schemes in total, comprising four control strategies for single synthetic jet actions and three for simultaneous multi-jet actions.
The innovative contributions are summarized as follows:

\begin{itemize}
    \item \textit{Sensitivity analysis of synthetic jet placement:}  
    Studies by \citeauthor{tangRobust2020} and \citeauthor{rabault2019artificial} highlight DRL's superior decision-making in effective flow control.
    \citeauthor{tangRobust2020} achieved a 5.7\% reduction in drag at $Re = 100$ using two pairs of synthetic jets. In contrast, \citeauthor{rabault2019artificial}, employing the same control algorithm under similar baseline flow conditions, achieved an 8\% drag reduction with only a pair of synthetic jets. 
    These findings underscore the critical impact of synthetic jet number and placement on control performance, warranting further investigation.
    Building on previous studies \cite{tangRobust2020,rabault2019artificial,yan2024aero,Suzuki2006,Ma2024}, we recognize that the placement of synthetic jets plays a pivotal role in determining control performance, directly influencing the effectiveness and energy efficiency of flow control strategies. 
    
    \item \textit{Comparison of single-jet and multi-jet control:} \citeauthor{yan2024aero} demonstrated in their study of flow around a square cylinder that multiple synthetic jet actuators outperform a single pair, highlighting the potential advantages of multi-jet configurations. However, this conclusion contrasts with findings from studies on circular cylinder flows \cite{tangRobust2020, rabault2019artificial}. To address this discrepancy, our study uniquely compares the performance of single-jet and multi-jet actuation in both square and circular cylinder flow scenarios. By exploring the synergistic effects of multi-jet actuation, we aim to uncover nonlinear flow interactions and provide insights into optimizing actuator design for balanced control efficiency and energy consumption.
    
    \item \textit{Achieving complete suppression of vortex shedding:} previous studies have achieved notable success in drag reduction; however, a common issue persists: the controlled flow in their results still exhibit vortex shedding, leading to oscillations in both lift and drag coefficients, even at $Re = 100$ \cite{rabault2019artificial, tangRobust2020, paris2021robust, Mao2022, he2023policy}. This observation prompted us to delve deeper into the underlying mechanisms. Drawing on vortex-induced vibration theory \cite{vortex}, it is reasonable to assume that the residual oscillations in lift and drag coefficients are directly linked to the incomplete suppression of vortex shedding in the controlled flow. 
    We tackle this challenge with a DRL-based active flow control strategy that completely eliminates vortex shedding and fully stabilizes the flow. Unlike previous studies reporting persistent instabilities, our results show lift and drag coefficients without any oscillatory behavior under identical conditions.

    \item \textit{Enhancing energy efficiency in flow control}: another significant contribution of this study is the exploration of energy-efficient control strategies. We observe that existing studies fail to comprehensively evaluate the control costs when assessing flow control performance. For active flow control strategies to be feasible, they must not only achieve the desired outcomes but also ensure energy efficiency. This requires the control strategies to optimize flow behavior with minimal energy input, thus preventing excessive energy consumption due to the control measures themselves. We not only ensures stable performance but also guarantees that the control solutions are optimized for minimal energy consumption.
\end{itemize}

The structure of this paper is organized as follows:  
In section \ref{sec:Methodology}, we introduce the research problem and detail the framework for integrating DRL with AFC.  
Section \ref{sec:cy} focuses on the DRL training results for circular cylinders, highlighting the optimization of jet placement and control strategies.  
Similarly, section \ref{sec:sq} presents the results for square cylinders, emphasizing the unique flow characteristics and control challenges.  
In section \ref{sec:Discussion}, we analyze the physical phenomena underlying the DRL-trained control strategies for both cylinder geometries, offering insights into the flow dynamics and the effects of jet configurations.  
Finally, section \ref{sec:Conclusions} summarizes the main findings of this study, discusses the implications for actuator design.

\section{METHODOLOGY}\label{sec:Methodology} 

DRL algorithms are employed to manipulate synthetic jets for active flow control, focusing on the effect of jet positioning on flow performance around square and circular cylinder at $Re = 100$. 
The numerical methods, software, and key parameters are outlined in section \ref{sec:Numerical simulation}. 
Section \ref{sec:Circular cylinder environment} describes the flow characteristics and jet arrangements for circular cylinders, while section \ref{sec:Square cylinder environment} provides similar details for square cylinders. 
Section \ref{sec:Deep Reinforcement Learning} introduces the core concepts, mathematical models, and distinctions of DRL, with an emphasis on the PPO algorithm. 
Finally, section \ref{sec:Enhanced Active Flow Control} demonstrates the application of DRL algorithms to address flow control challenges.

\subsection{Numerical simulation}\label{sec:Numerical simulation}

The behavior of incompressible viscous fluid flow within the domain $\Omega \subset \mathbb{R}^{nd}$ over the time interval $(0, T)$ is described using the Navier-Stokes equations. These equations govern the evolution of the fluid velocity field $\mathbf{u} = \mathbf{u}(\mathbf{x}, t)$ and pressure field $p = p(\mathbf{x}, t)$, where $\mathbf{x}$ represents the spatial coordinates and $t$ denotes time.
The governing equations in their non-dimensional form are expressed as:
\begin{subequations}
\begin{equation}
    \frac{\partial \mathbf{u}}{\partial t} + \mathbf{u} \cdot (\nabla \mathbf{u}) = -\nabla p + Re^{-1} \Delta \mathbf{u} \quad \text{in} \quad \Omega \times (0, T),
\end{equation}
\begin{equation}
    \nabla \cdot \mathbf{u} = 0 \quad \text{in} \quad \Omega \times (0, T),
\end{equation}
\end{subequations}
where the Reynolds number is defined as $Re = \frac{UD}{\nu}$, with $U$ being the upstream inlet velocity, $D$  as the characteristic length (the cylinder diameter for a circular cylinder and the length of one side for a square cylinder), and $\nu$ denoting the kinematic viscosity of the fluid. 

We employ \texttt{OpenFOAM}, a widely used Navier-Stokes solver recognized in both industry and academia \cite{jasakLibrary}, for numerical simulations. The computational domain is discretized using the finite volume method to capture complex fluid behavior. The \texttt{pimpleFoam} solver, combining the \texttt{SIMPLE} and \texttt{PISO} algorithms \cite{ISSA198640,SIMPLE}, efficiently resolves unsteady Navier-Stokes equations, enabling larger time steps with stability and computational efficiency.

\subsection{Circular cylinder environment}\label{sec:Circular cylinder environment}

\begin{figure*}[t]
    \centering
    \includegraphics{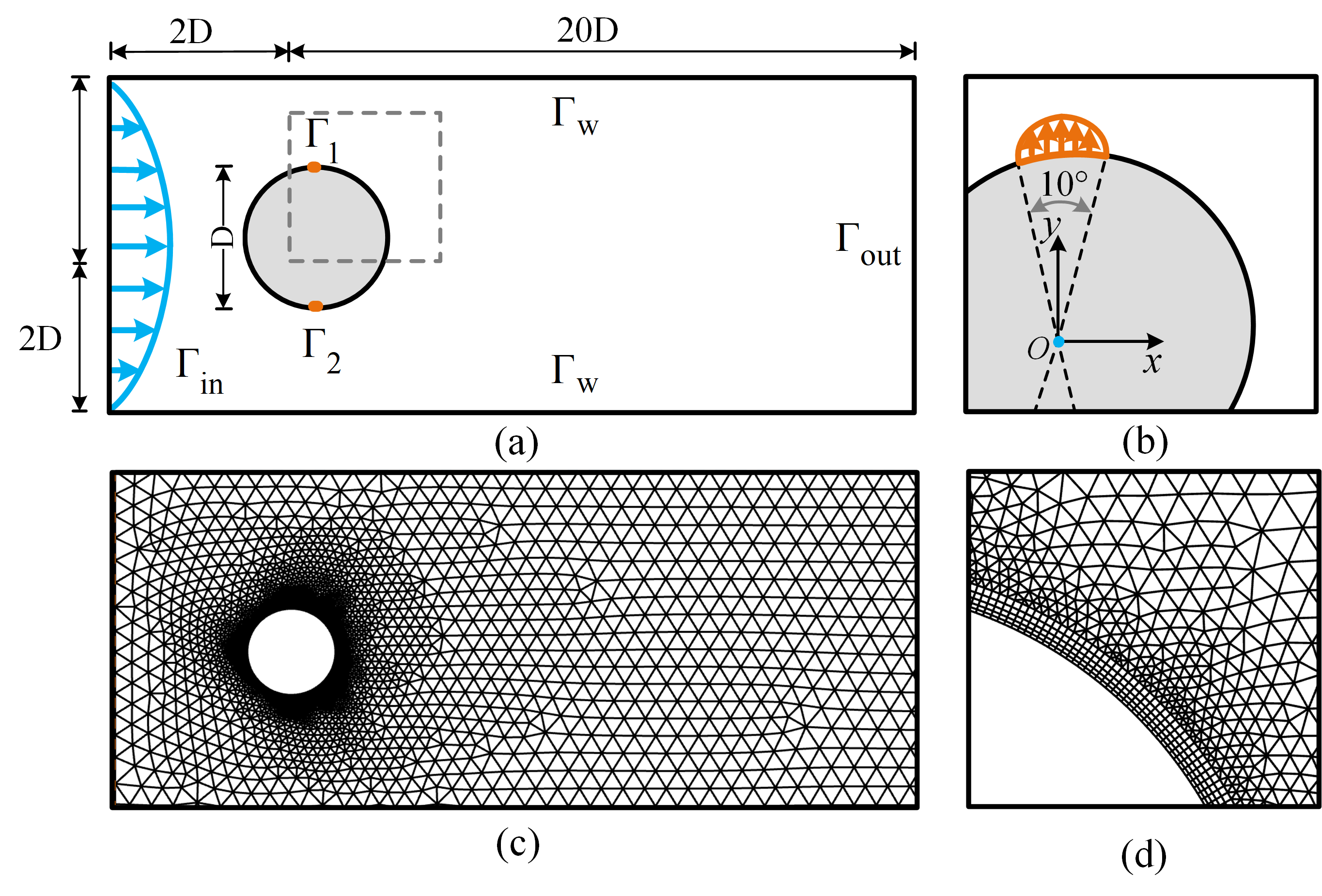}
    \caption{Schematics of the physical model of flow around a circle cylinder.
    (a) Domain configuration and boundary conditions.
    (b) Details of the jets conditions on the cylinder.
    (c) Mesh structure around the cylinder, and (d) Zoomed-in view of the mesh near the cylinder boundary.}
\label{fig:figure1}
\end{figure*}

The configuration used for simulating flow around a circular cylinder closely follows the classical benchmark established by \citeauthor{Schafer1996} 
As illustrated in figure \ref{fig:figure1}, the cylinder is placed within a rectangular computational domain with dimensions of $22D$ along the $x$-axis and $4.1D$ along the $y$-axis. 
The origin of the Cartesian coordinate system is positioned at the center of the cylinder. 
The cylinder is slightly offset in the $y$-direction to induce vortex shedding, facilitating the analysis of oscillatory flow phenomena.
The flow is characterized by a kinematic viscosity of \(\nu = 0.01 \, \text{m}^2/\text{s}\), with a characteristic length \(L = 1 \, \text{m}\) and velocity \(U = 1 \, \text{m/s}\), corresponding to a Reynolds number of 100.
The boundaries of the computational domain are divided into an inlet $\Gamma_\text{in}$, an outlet $\Gamma_\text{out}$, no-slip walls $\Gamma_\text{w}$, and two separate jets on the cylinder $\Gamma_{i}\, (i = 1,2)$, as shown in~figure \ref{fig:figure1}(a). 
At the inlet $\Gamma_\text{in}$, the inflow velocity along $x$-axis is prescribed by a parabolic velocity profile in the form,
\begin{equation}
U_{\text{inlet}}(y) = U_m \left( (H - 2y)(H + 2y) / H^2 \right),
\end{equation}
and that along $y$-axis is prescribed as $V_{\text{inlet}}(y) = 0$.
$U_m$ is the maximum velocity magnitude of the parabolic profile, and $H=4.1D$ represents the total height of the rectangular domain. The average inlet velocity $\overline{U}$, is related to the parabolic velocity profile $U_{\text{inlet}}(y)$ through the expression:
\begin{equation}
\overline{U}=\frac{1}{H} \int_{-H/2}^{H/2} U_{\text{inlet}}(y) d y=\frac{2}{3} U_m.
\end{equation}

Additionally, a no-slip boundary condition is enforced on the areas of the cylinder surface where the jets are not located.
At the outlet boundary $\Gamma_\text{out}$, an outflow condition is applied, where the velocity is extrapolated based on the internal flow field. 
No-slip boundary conditions are also applied at the upper and lower walls $\Gamma_\text{w}$. 
This specification of boundary conditions ensures accurate flow simulation and effective control through synthetic jets.
Mathematically, the boundary conditions are written as:
\begin{eqnarray}
\begin{array}{cl}
-\rho \mathbf{n} \cdot \mathbf{p} + Re^{-1} (\mathbf{n} \cdot \nabla \mathbf{u}) = 0  &\text{on\quad } \;\Gamma_{out}, \\
\mathbf{u} = 0  &\text{on \quad} \Gamma_{w}, \\
\;\mathbf{u} = U  &\text{on \quad} \Gamma_{in}, \\
\;\;\;\mathbf{u} = f_{Q_i}  & \text{on \quad} \Gamma_{i}, \quad i = 1, 2.
\end{array}
\end{eqnarray}
where $\mathbf{n}$ denotes the normal vector and $f_{Q_i}$ represents the radial velocity profiles simulating the suction or injection of fluid by the jets.
Flow control is implemented through synthetic jet pairs positioned at the cylinder, as illustrated in~figure \ref{fig:figure1}(b). These synthetic jet pairs, denoted as $\Gamma_i$ $(i = 1,2)$, alternately inject and suction fluid to actively manipulate the surrounding flow field.

\begin{figure*}[t]
    \centering
    \includegraphics{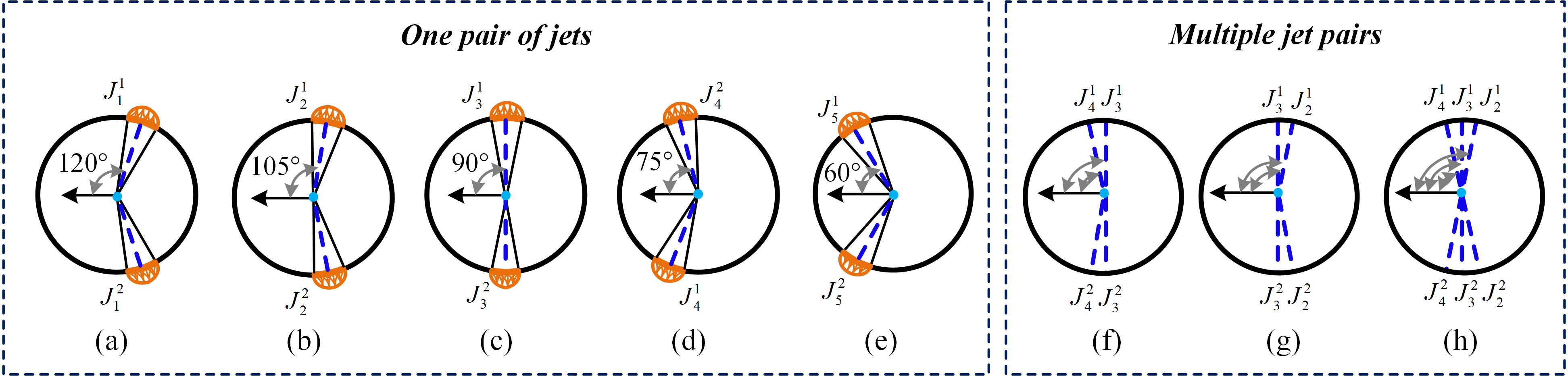}
    \caption{ 
    For a pair of jets, five arrangements are considered: 
    (a) $J_1^1$ and $J_1^2$, 
    (b) $J_2^1$ and $J_2^2$, 
    (c) $J_3^1$ and $J_3^2$, 
    (d) $J_4^1$ and $J_4^2$, 
    (e) $J_5^1$ and $J_5^2$. 
    For two pairs of jets, two arrangements are examined: 
    (f) $J_3^1$, $J_3^2$ and $J_4^1$, $J_4^2$; 
    (g) $J_3^1$, $J_3^2$ and $J_2^1$, $J_2^2$. 
    For three pairs of jets, one arrangement is studied: 
    (h) $J_2^1$, $J_2^2$, $J_3^1$, $J_3^2$, and $J_4^1$, $J_4^2$.}
    \label{fig:figure2}
\end{figure*}

In the numerical simulation, the synthetic jets are modeled with a parabolic velocity distribution, where the velocity peaks at the jet center and decreases to zero at the edges. Each jet spans an angular width of $\omega = 10^{\circ}$ and is oriented perpendicularly to the cylinder surface. The control action directly adjusts the amplitude of this parabolic profile, thereby regulating the jet intensity.  
A positive value of the control action represents that the synthetic jet is in a blowing state, and a negative value of the control action represents that the synthetic jet is in a suction state.
To maintain global mass conservation, the net mass flow rate of the jet pair is constrained such that $V_{\Gamma_1} = -V_{\Gamma_2}$, ensuring that no net mass is introduced or removed from the system.  
Through this mechanism, synthetic jet pairs dynamically influence the near-wall flow characteristics, providing an effective means to modulate vortex shedding, reduce drag, and enhance wake stability.

To systematically investigate the influence of synthetic jet positioning on flow control around a circular cylinder, experiments are conducted using three configurations: a single-pair, two-pairs and three-pairs jet arrangement. The specific placements of the jets in each configuration are detailed in figure \ref{fig:figure2}.  
Through this mechanism, synthetic jet pairs dynamically influence the near-wall flow characteristics, providing an effective means to modulate vortex shedding, reduce drag, and enhance wake stability.

\begin{itemize}
    \item \textit{Single-Pair Configuration:} Five distinct jet placement schemes are evaluated, where the jet pair is denoted as $ J_1^1 $ and $ J_1^2 $. The subscript represents the pair index, while the superscript distinguishes individual jets within the pair. The jets are positioned at azimuthal angles of $60^\circ$, $75^\circ$, $90^\circ$, $105^\circ$, and $120^\circ$, enabling an assessment of the impact of individual jet placement on flow control effectiveness.
    
    \item \textit{Two-pairs Configuration:} Two placement schemes are considered to analyze the interaction effects between jet pairs. In this setup, additional jets are introduced to examine how multiple jets influence vortex dynamics and flow stabilization compared to the single-pair case.
    
    \item \textit{Three-Pairs Configuration:} This configuration activates six jets, designated as $ J_2^1 $, $ J_2^2 $, $ J_3^1 $, $ J_3^2 $, $ J_4^1 $, and $ J_4^2 $. The objective is to evaluate the cumulative effect of multiple jet pairs on flow control and identify potential synergies that enhance performance.
\end{itemize}

\subsection{Square cylinder environment}\label{sec:Square cylinder environment}

\begin{figure*}[h]
    \centering
    \includegraphics{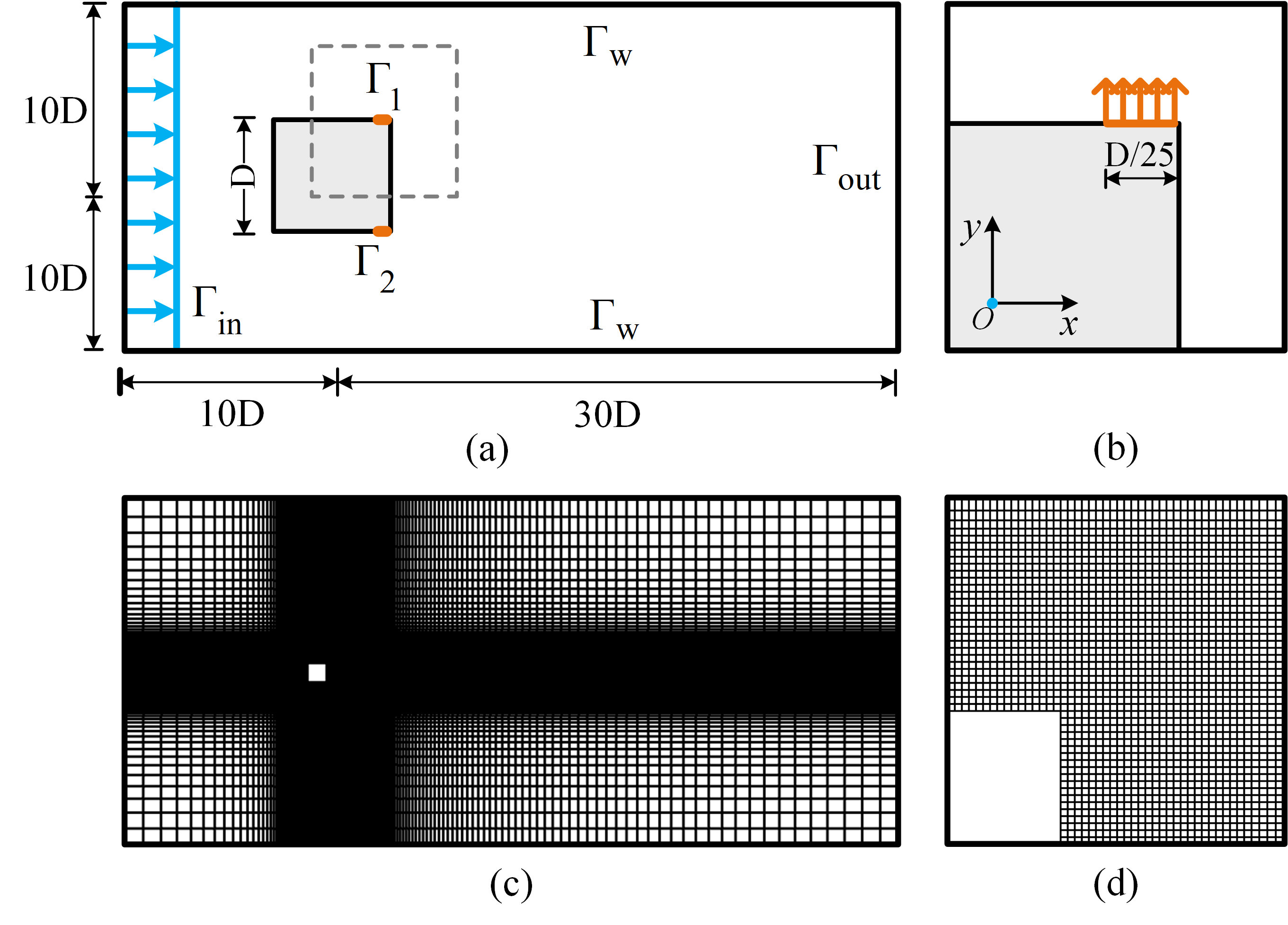}
    \caption{Schematic of the computational domain and boundary conditions. (a) Establishment of the coordinate system and detailed dimensions of the computational domain. (b) Placement of the synthetic jets near the trailing corner point. (c) The mesh generation scheme around the square object, illustrating the distribution of high-density mesh in regions prone to flow separation, and (d) A close-up view of the mesh structure.}
\label{fig:figure3}
\end{figure*}

The computational domain for simulating the flow around a square cylinder is defined as a rectangle with dimensions $ 40D \times 20D $, where $ D $ denotes the side length of the square cylinder. A Cartesian coordinate system is established with the center of the square cylinder as the origin, and the positive $ x $-axis represents the streamwise direction. The domain extends $ 10D $ upstream and $ 30D $ downstream from the cylinder center, while the vertical extent spans $ \pm 10D $. 
As illustrated in figure \ref{fig:figure3}(a), the computational domain and boundary conditions are configured to minimize boundary effects and accurately capture wake dynamics. A uniform velocity profile is imposed at the inlet boundary $ \Gamma_{\text{in}} $, corresponding to a Reynolds number of 100. The velocity of the synthetic jets is directed normal to the cylinder walls, as depicted in figure \ref{fig:figure3}(b). A no-slip boundary condition is applied to the remaining surfaces of the square cylinder. At the outlet boundary $ \Gamma_{\text{out}} $, a Neumann condition is enforced to maintain a zero-stress vector, ensuring a physically consistent outflow. Additionally, far-field conditions are imposed on the upper and lower boundaries to minimize their influence on the flow around the cylinder.

To ensure numerical stability, a time step of $ \Delta t = 0.0005 $ is selected based on extensive validation, striking a balance between temporal accuracy and compliance with the CFL condition. This choice minimizes numerical errors and enhances the fidelity of the simulation. 
The grid independence study for the numerical simulation of flow around the square cylinder is presented in Appendix \ref{app1}.
The computational mesh used in subsequent simulations, shown in figure \ref{fig:figure3}(c), gradually coarsens radially outward from the square cylinder to optimize computational efficiency. 
Meanwhile, figure \ref{fig:figure3}(d) highlights local grid refinement near the cylinder.

\begin{figure*}[htb!]
    \centering
    \includegraphics{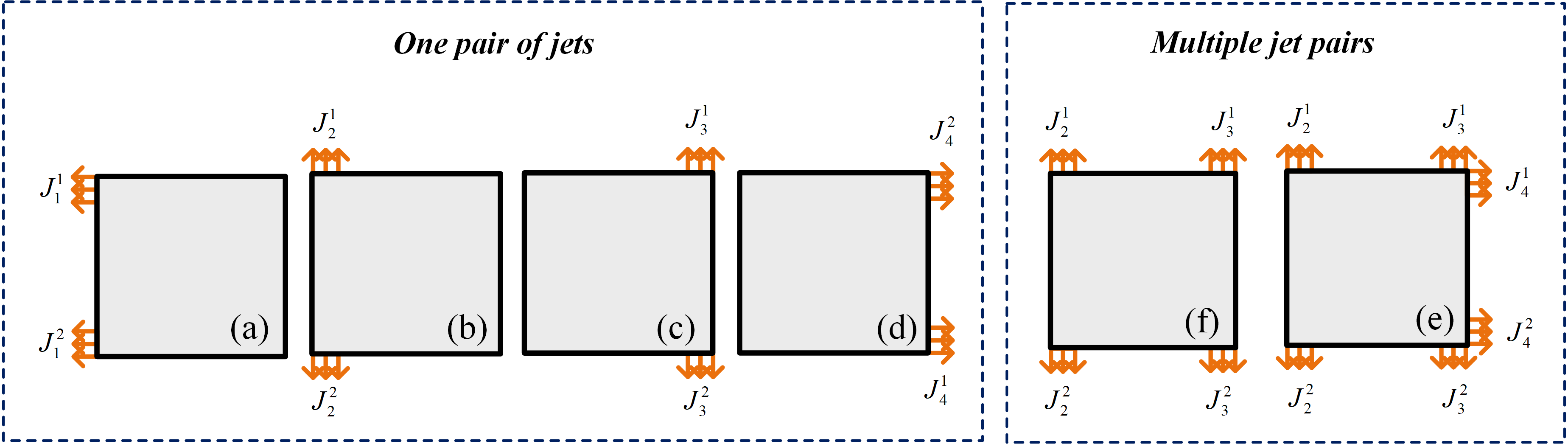}
    \caption{Jets arrangement schemes for the flow around a square cylinder. 
    For a single pair of jets, five different arrangements are considered: 
    (a) $J_1^1$ and $J_1^2$;
    (b) $J_2^1$ and $J_2^2$;
    (c) $J_3^1$ and $J_3^2$;
    (d) $J_4^1$ and $J_4^2$.
    For two pairs of jets, two arrangements are examined:  
    (e) $J_2^1$, $J_2^2$, $J_3^1$ and $J_3^2$; 
    For three pairs of jets, one arrangement is studied: 
    (f) $J_2^1$, $J_2^2$, $J_3^1$, $J_3^2$, $J_4^1$ and $J_4^2$.}
\label{fig:figure4}
\end{figure*}

To investigate the sensitivity of control performance to synthetic jet placement and the synergistic effects of multiple jet pairs, we design configurations where one, two, or three pairs of synthetic jets are activated simultaneously.  
For the single-pair jet configuration, four distinct placement schemes are considered. Each jet is named according to its position, with $ J_1^1 $ and $ J_1^2 $ representing jets located near the front corners. Additionally, configurations are designed where two pairs and three pairs of jets operate simultaneously.  
In multi-pair jet configurations, the DRL algorithm assigns independent control actions to each jet pair, allowing for a detailed evaluation of how different placements affect flow control effectiveness. The various jet placement schemes are illustrated in figure \ref{fig:figure4}.

\subsection{Deep Reinforcement Learning}\label{sec:Deep Reinforcement Learning}

Deep learning, a branch of machine learning, utilizes deep neural networks with input, hidden, and output layers to process raw data, extract features, and generate task-specific predictions \cite{lecun2015deep, granter2017alphago}. 
Reinforcement learning complements this by enabling agents to learn optimal actions in dynamic environments through trial and error, refining strategies by maximizing cumulative rewards \cite{mnih2013playing, 9904958}.

Deep learning is a branch of machine learning that uses deep neural networks with input, hidden, and output layers \cite{lecun2015deep}.
These networks process raw data to extract relevant features and generate predictions for specific tasks \cite{granter2017alphago}.
Reinforcement Learning is a machine learning paradigm where an agent learns optimal actions in a dynamic environment through trial and error \cite{mnih2013playing}. By maximizing cumulative rewards via feedback, the agent iteratively refines its strategy to adapt to environmental complexities and make effective decisions \cite{9904958}.

\subsubsection{Foundation}
As illustrated in figure \ref{fig:figure5}(a), the agent interacts with the environment by taking actions ($a \in \mathbb{A}$) based on its policy ($\pi(a|s)$), which guides action selection from the current state ($s \in \mathbb{S}$) \cite{franccois2018introduction}. The environment provides feedback as rewards, enabling the agent to optimize its strategy. The state value function ($V(s)$) estimates the expected cumulative reward from a state, while the action value function ($Q(s, a)$) evaluates the expected reward from a specific action in a given state \cite{9904958}.

\subsubsection{Mathematical model}
RL relies on a finite Markov Decision Process (MDP), defined by the tuple $ (S, A, P, R, \gamma) $, where $ S $ is the state space, $ A $ the action space, $ P $ the state transition probability, $ R $ the reward function, and $ \gamma $ the discount factor \cite{PUTERMAN1990331,altman1999}. At each time step $ i $, the agent observes state $ s_i $, selects action $ a_i $, transitions to state $ s_{i+1} $ with probability $ p(s_{i+1} | s_i, a_i) $, and receives reward $ R(s_i, a_i, s_{i+1}) $. Decision-making is guided by a policy $ \pi(a_i | s_i) $, defining the probability of choosing $ a_i $ in state $ s_i $.

\subsubsection{Classification}

\begin{figure*}[t]
    \centering
    \includegraphics[width=0.8\textwidth]{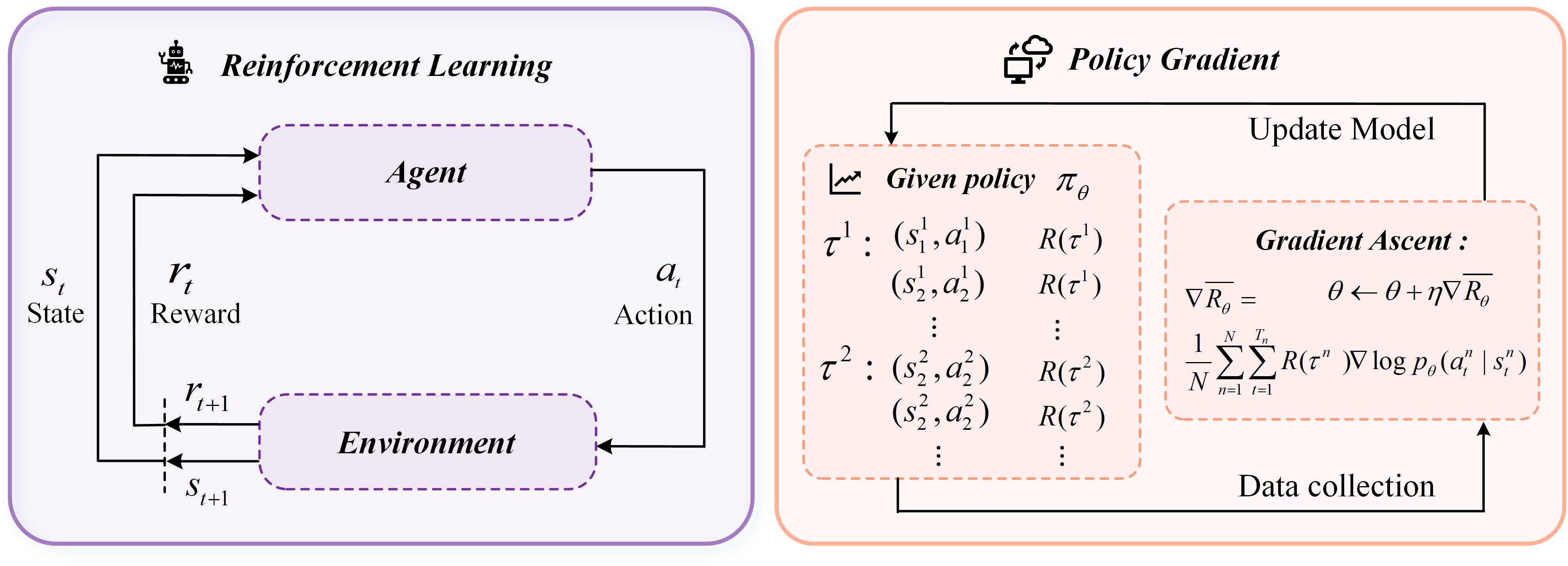}
    \caption{Basic structures of (a) Reinforcement Learning and (b) Policy gradient.}
\label{fig:figure5}
\end{figure*}

In reinforcement learning, the value function $ V(s) $ and the action-value function $ Q(s, a) $ serve as critical metrics for assessing the potential of states and state-action pairs to yield high cumulative rewards.
These are defined as follows \cite{szepesvari}:
\begin{align}
\begin{aligned}
V(s) &= \mathbb{E}_{\pi} \left[ \sum_{t=0}^{\infty} \gamma^t r_t \mid s_0 = s \right], \\
Q(s, a) &= \mathbb{E}_{\pi} \left[ \sum_{t=0}^{\infty} \gamma^t r_t \mid s_0 = s, a_0 = a \right],
\end{aligned}
\end{align}
where $ \mathbb{E}_{\pi} $ denotes the expectation under the policy $ \pi $, $ \gamma $ is the discount factor ($ 0 \leq \gamma \leq 1 $), and $ r_t $ is the immediate reward at time step $ t $. $ s_0 = s $ indicates that the initial state is $ s $, and $ a_0 = a $ specifies the initial action for $ Q(s, a) $.  
These functions collectively describe the expected cumulative rewards for states and state-action pairs, guiding the agent in optimizing its policy to maximize long-term returns.

RL is broadly categorized into value-based and policy-based methods. Value-based methods, such as Q-learning and DQN, learn $ V(s) $ or $ Q(s, a) $ to estimate cumulative rewards and select optimal actions by maximizing these values \cite{mnih2015human,altman1999}. Policy-based methods directly optimize a policy function ($ \pi(a|s) $) to maximize rewards, excelling in continuous action spaces and stochastic policies. Examples include policy gradient methods, TRPO, and PPO \cite{PG,TRPO,schulman2017}. The structure of policy-based training is shown in figure \ref{fig:figure5}(b).

Reinforcement learning is broadly categorized into value-based and policy-based methods. Value-based methods, such as Q-learning and Deep Q-Networks, learn value functions like $ V(s) $ or $ Q(s, a) $ to estimate cumulative rewards. Optimal actions are selected by maximizing these values, making value-based approaches particularly effective in discrete action spaces \cite{mnih2015human, altman1999}.  
Policy-based methods take a different approach by directly optimizing a policy function $ \pi(a|s) $ to maximize expected rewards. These methods are well-suited for problems involving continuous action spaces and stochastic policies. Policy gradient methods, including Trust Region Policy Optimization and Proximal Policy Optimization, refine policies through iterative updates, ensuring stability and efficiency in training \cite{PG, TRPO, schulman2017}.  Figure \ref{fig:figure5}(b) illustrates the structure of policy-based training, highlighting its iterative nature and optimization process.

\subsubsection{policy optimization algorithm}

PPO is a robust policy optimization algorithm that stabilizes performance by constraining policy update amplitudes. It uses a clipped objective to keep the probability ratio of the current and old policies within a safe range, preventing performance fluctuations. By leveraging the advantage function and optimizing via stochastic gradient descent, PPO efficiently learns the optimal policy \cite{schulman2017}.

\begin{equation}
L^{\text{CLIP}}(\theta) = \hat{\mathbb{E}}_t \left[ \min \left( r_t(\theta) \hat{A}_t, \text{clip}(r_t(\theta), 1 - \epsilon, 1 + \epsilon) \hat{A}_t \right) \right],
\end{equation}
where $ r_t(\theta) = \frac{\pi_\theta(a_t | s_t)}{\pi_{\theta_\text{old}}(a_t | s_t)} $ measures policy change, and the advantage estimate $ \hat{A}_t $ evaluates the benefit of action $ a_t $ in state $ s_t $. A clipping mechanism, $ \text{clip}(r_t(\theta), 1 - \epsilon, 1 + \epsilon) $, ensures stability by restricting policy updates within $ [1 - \epsilon, 1 + \epsilon] $, where $ \epsilon $ controls the update range. 

\subsection{Enhanced Active Flow Control}\label{sec:Enhanced Active Flow Control}

Through a DRL-based AFC framework, we integrate a CFD environment with DRL algorithms to address flow control problems.
The PPO agent is employed to train a control policy that optimizes real-time actuation of two synthetic jets positioned on the cylinder's surface, with the objective of minimizing fluid forces. The agent perceives the CFD environment through strategically placed flow probes that measure velocity and pressure. Based on these observations, the agent dynamically adjusts the mass flow rate of the synthetic jets to manipulate the flow field. This framework, comprising the agent, environment, state observations, action selection, and reward mechanism, is meticulously detailed to elucidate the intricate interplay among these components in enhancing the performance of AFC.

\begin{figure*}
    \centering
    \includegraphics[width=0.85\textwidth]{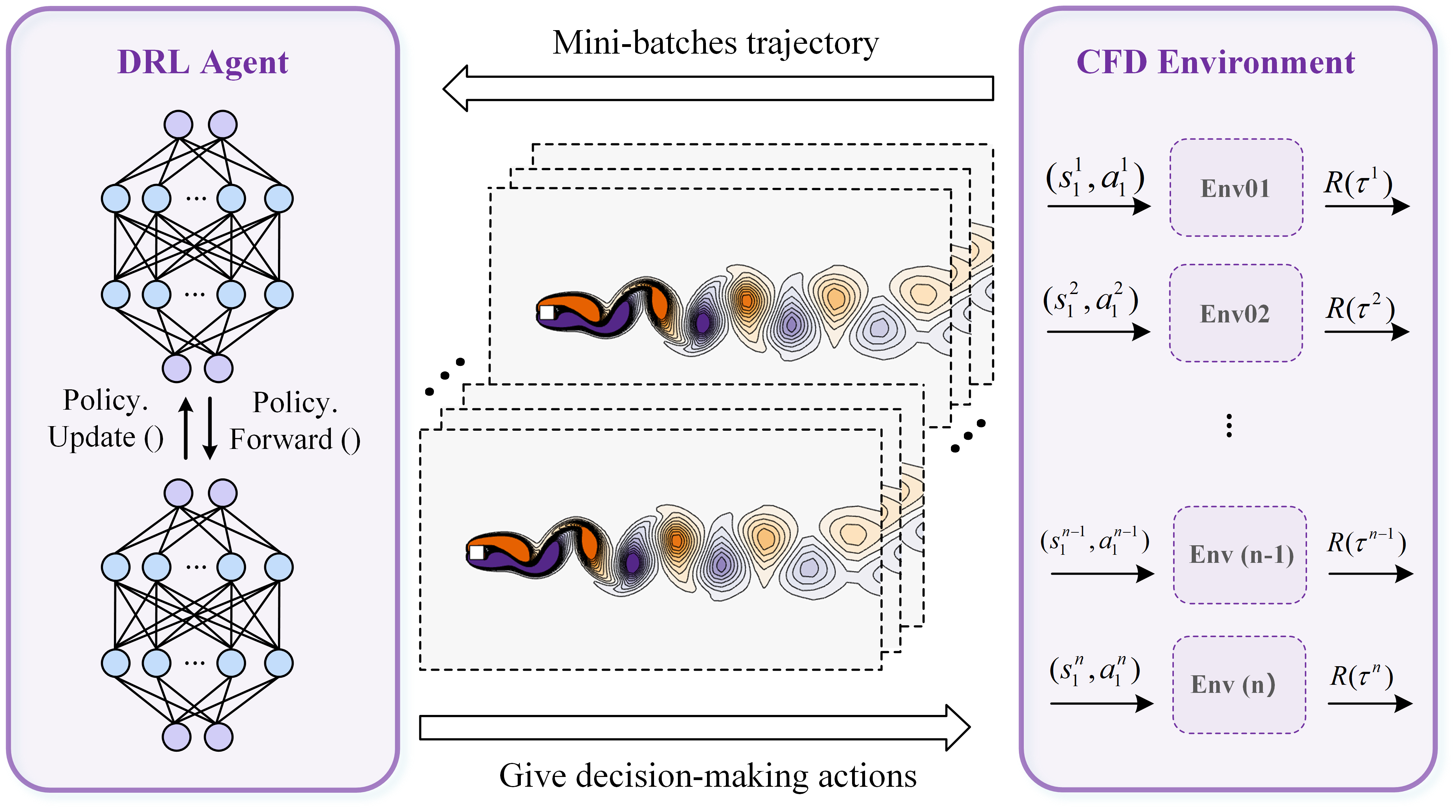}
      \caption{Schematic of the interaction between the deep reinforcement learning agent and the computational fluid dynamics environment for active flow control. The DRL agent receives state information $ s_t^i $ from the CFD environment at each timestep $ t $, selects an action $ a_t^i $ to adjust the synthetic jets, and then the environment returns a new state $ s_{t+1}^i $ and a reward $ R(\tau^i) $, where $ \tau^i $ represents the trajectory of states and actions. The agent updates its policy $\pi_\theta$ to maximize the expected cumulative reward $ \mathbb{E}[R(\tau^i)] $, optimizing flow control through iterative learning across multiple parallel environments.}
\label{fig:fig6}
\end{figure*}

\begin{itemize}
    \item Agent : agent selection depends on the state space, action space, exploration-exploitation trade-off, computational resources, and objectives. For high-dimensional continuous spaces like fluid mechanics, PPO \cite{rabault2019artificial,rabault2020deep,tangRobust2020}, Soft Actor-Critic (SAC) \cite{wangDRLinFluids,jia2024robust,Xia2024}, Twin Delayed Deep Deterministic Policy Gradient (TD3) \cite{Dixiapnas} and Truncated Quantile Critics (TQC) \cite{Xia2024} are ideal, with PPO excelling in real-time flow control due to its exploration balance and efficiency.
    \item Environment : the environment interacting with the agent is instantiated using CFD.
    Figure \ref{fig:fig6} illustrates the integration of CFD computations as a DRL training environment and demonstrates the benefits of accelerating the training process by using multiple parallel CFD environments.
    This parallelization strategy is inspired by the work of \cite{rabaultAccelerating}, and also incorporates insights from \cite{jia2024optimal}. 
    The environment depends on the type of CFD solver, and there are various codes available for coupling numerical simulation environments with RL frameworks. 
    RL has been integrated with the \texttt{FEniCS} environment \cite{rabault2019artificial}, 
    the \texttt{OpenFOAM} environment \cite{wangDRLinFluids}, 
    and the \texttt{Nek5000} environment \cite{liReinforcementlearning}.

    \item Action $a_t$: the agent's actions are the dimensionless mass flow rates of the synthetic jets, $ Q_i $ ($ i = 1, 2 $), constrained by $ Q_1 + Q_2 = 0 $ to ensure mass conservation and improve numerical stability.
    The control actions for the synthetic jets are defined such that a positive value indicates the jet is in the blowing state, while a negative value indicates the jet is in the suction state. To ensure the stability of the system and prevent non-physical large actuations that could destabilize it, a limit is imposed on the normalized mass flow rate of the jets. Additionally, to further enhance stability and mitigate any non-physical instabilities in the flow, a smoothing function is applied to the control actions.

    \item State $s_t$ : 
    the agent observes the CFD environment by collecting physical field data from strategically placed probes. Inspired by \cite{wang2024dynamic}, \cite{rabault2019artificial}, and \cite{liReinforcementlearning}, probes are positioned in wake regions with high instability and sensitivity, optimizing flow control by monitoring time-averaged fluctuating pressure. 
    Additionally, multiple sensor probes are distributed around the cylinder to capture detailed physical information. 
    \citeauthor{paris2021robust}, \citeauthor{liReinforcementlearning} and \citeauthor{wang2024dynamic} studied the impact of the distribution of observation probes on control performance.

    \item Reward $ r_t $: the reward function is designed to optimize flow control by reducing both drag and lift forces around the cylinder. 
    To facilitate comparison with prior research, our reward function does not incorporate external energy consumption \cite{rabault2019artificial,jia2024optimal,wangDRLinFluids,jia2024robust}. We aim to train the agent to develop a low-energy control strategy through its interaction with the environment.
    It is formulated as:
    \begin{equation}
    r_{T_i} = C_{D,0} - \left(C_D\right)_{T_i} - \omega\left|\left(C_L\right)_{T_i}\right|,
    \end{equation}
    where $ C_{D,0} $ represents the baseline drag coefficient, providing a reference for potential reward improvement. The  $ \left(\cdot\right)_{T_i} $ represents the sliding average over the duration of one jet flow control period $ {T_i} $. The parameter $ \omega $ is a weighting factor that balances the importance of reducing lift relative to drag, ensuring that the reward function effectively incentivizes the agent to minimize both forces simultaneously.
    
    \item Interaction: the agent interacts with the environment at each time step $ t $ by observing the current flow state $ s_t $, selecting an action $ a_t $ based on the policy $ \pi_{\theta}(a_t \mid s_t) $, and receiving a reward $ r_t $ and the next state $ s_{t+1} $. The agent iteratively refines $ \pi_{\theta} $ to maximize cumulative rewards $ \sum_{t} \gamma^t r_t $. 
    The DRL training process encompasses 3,000 episodes, each comprising 100 timesteps. Each timestep corresponds to 0.025 non-dimensional time units, encompassing approximately 8 to 10 vortex shedding cycles. The PPO algorithm iteratively updates the policy using mini-batch optimization across all 3,000 episodes.
\end{itemize}

\section{Results: flow control performance of a circular cylinder}\label{sec:cy}

This section examines the impact of synthetic jet positioning on flow control performance and costs, quantified by dimensionless mass flow rates. Five single-pair jet configurations, two two-pairs configurations, and one additional single-pair configuration are analyzed. 
It is important to note that during the training process, only the positions or quantities of the synthetic jets are altered, while the hyperparameters for each DRL training remain consistent.

In our DRL framework, a single agent is designed to control multiple actions corresponding to the manipulation of multiple synthetic jets. This unified control approach ensures that the agent optimizes its policy considering the overall system dynamics, rather than relying on cooperation among multiple agents. The agent receives a global state representation and outputs actions that collectively influence the environment, allowing for efficient and coordinated control of all jets. Importantly, the agent operates based on a single reward function that evaluates the overall performance of the system, ensuring that the control strategy is optimized for the collective impact of all jets.

\begin{figure*}[htb!]
    \centering
    \includegraphics[width=\textwidth]{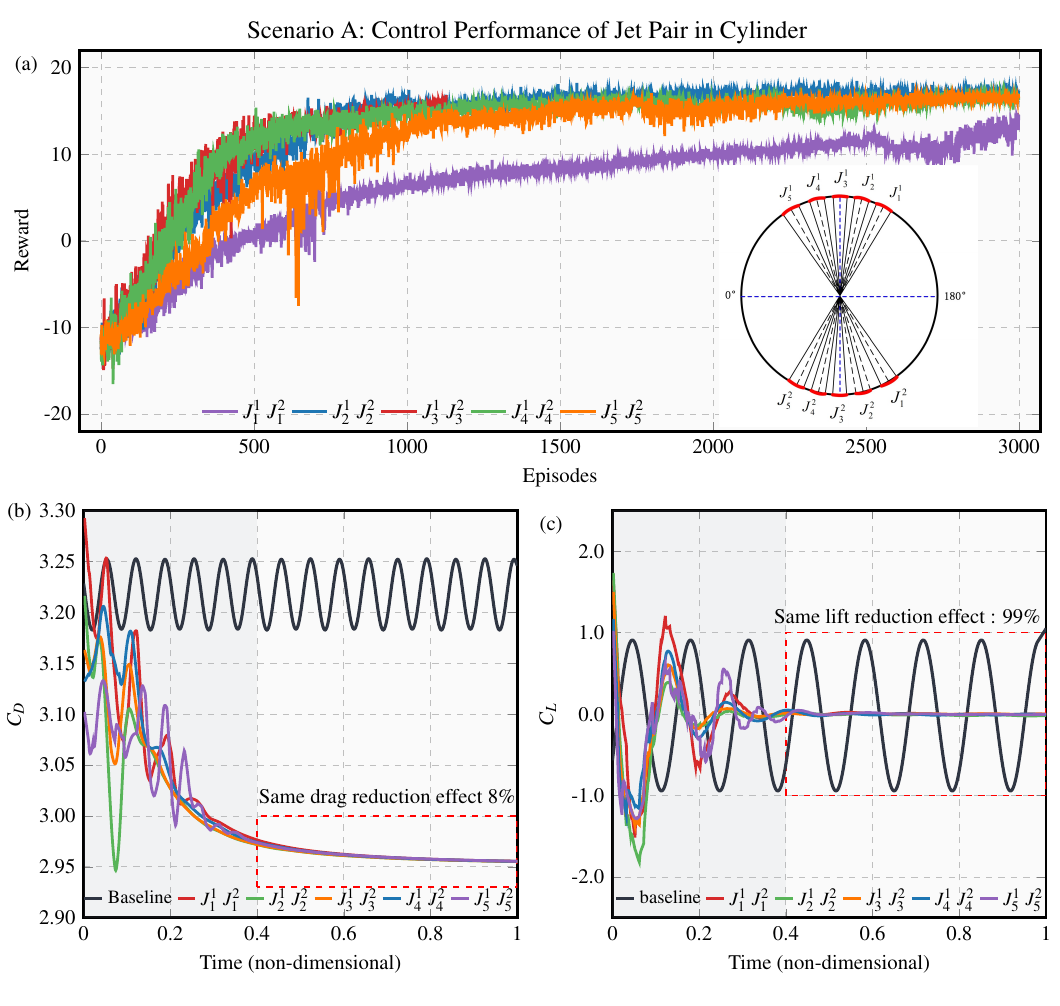}
    \caption{Evaluation of flow control performance for various jet configurations under the condition of activating only a single pair of synthetic jets. 
    (a) Reward curves over training episodes. 
    (b) The time history of the $C_D$ of the cylinder in both baseline and controlled flow conditions.
    (c) The time history of the $C_L$ of the cylinder in both baseline and controlled flow conditions.
    }
\label{fig:fig7}
\end{figure*}

\subsection{A pair of synthetic jets controllers}

\subsubsection{Scenario A: Control Performance of Jet Pair }

In this section, we focus solely on the sensitivity of DRL training control performance to the angular positions of single pairs of synthetic jets on the cylinder surface. Specifically, we compare and analyze the reward function curves, flow control strategies, and flow control performance for synthetic jets at different positions. The range of angular positions on the cylinder surface spans from 60° to 120°. For instance, the jet configuration $ J_1 $ comprises a pair of synthetic jets, $ J_1^1 $ and $ J_1^2 $, with $ J_1^1 $ located at an angle of 120°. Similarly, the $ J_5 $ configuration includes two jets, $ J_5^1 $ and $ J_5^2 $, with $ J_5^1 $ positioned at 60°. The positions of the remaining synthetic jets are designed in a similar manner.

The reward function curves for different jet positions are illustrated in Figure \ref{fig:fig7} when only a single pair of synthetic jets is activated. These curves reflect the performance of the DRL agent during training. The initial fluctuations indicate the exploration phase, while the subsequent upward trend signifies the gradual optimization of the control strategy. Comparing the reward function curves for different jet configurations, we find that the jets located at position $ J_1 $ show the slowest growth and the lowest final reward, highlighting the challenges in control. The jets at position $ J_5 $ demonstrate moderate convergence performance, whereas the jets at positions $ J_2 $, $ J_3 $, and $ J_4 $ exhibit rapid convergence. After 1500 episodes, the reward curves stabilize, indicating that the DRL agent has found a stable and effective control strategy capable of robustly reducing $ C_D $ and controlling $ C_L $.

To clearly demonstrate the control performance based on DRL, the drag and lift coefficients for both the baseline and controlled flows are presented in figures \ref{fig:fig7}(b) and \ref{fig:fig7}(c). 
The baseline flow, devoid of active flow control, serves as a reference. 
In stark contrast, the controlled flow, modulated by the DRL algorithm, consistently achieves an 8\% reduction in drag coefficient across all jet configurations. This reduction is not merely transient but is sustained over time, as evidenced by the stabilization of the drag and lift coefficients to a steady state following the initial transient phase. This suppression of periodic fluctuations underscores the robustness and effectiveness of the DRL-driven control mechanism in mitigating flow-induced instabilities.

\begin{figure*}[htb!]
    \centering
    \includegraphics[width=0.9\textwidth]{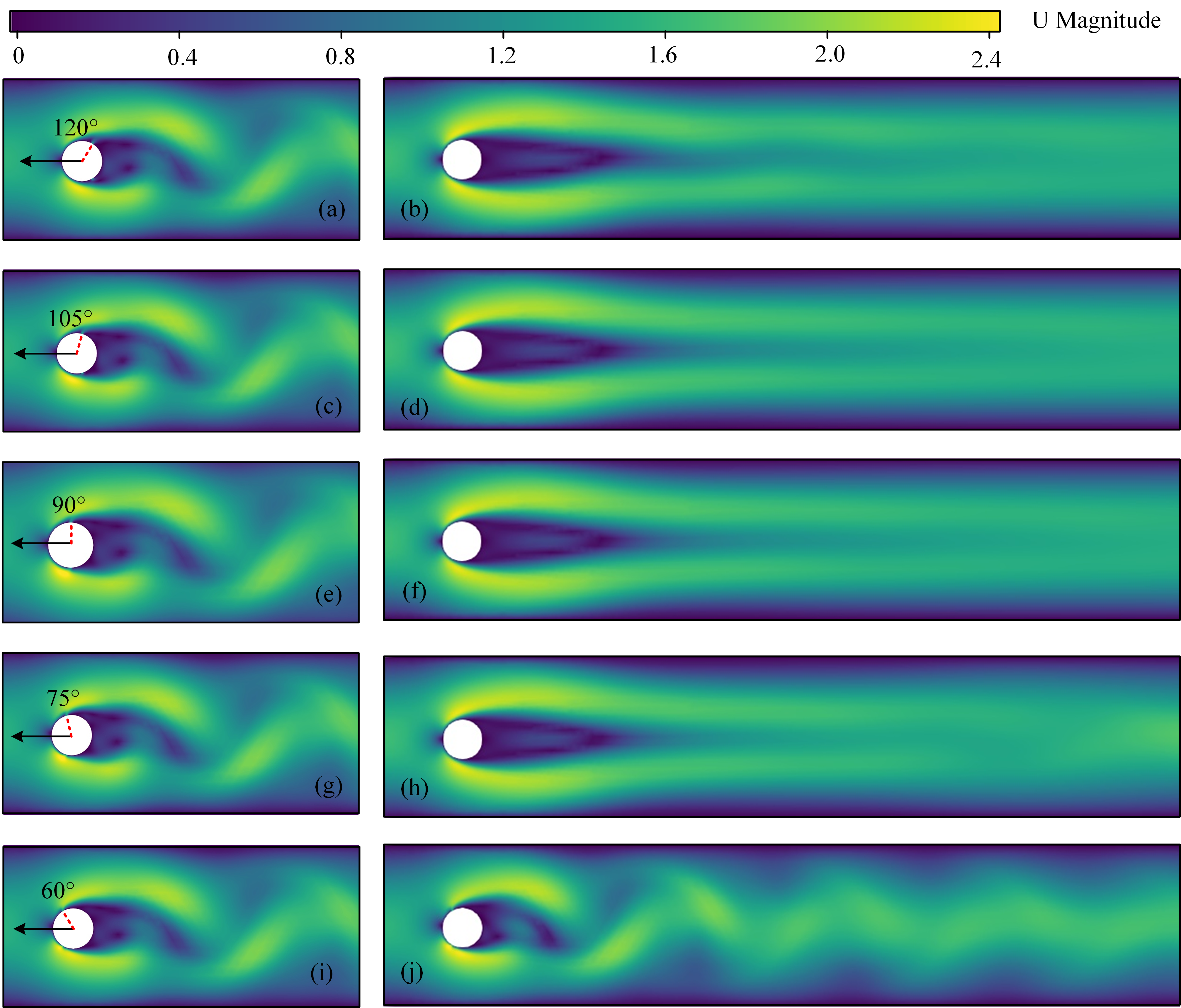}
    \caption{ Velocity contours of flow around a cylinder with synthetic jets at various angular positions. The left column shows the jet activation, and the right column shows the corresponding control results.
    (a) and (b) Jets at 120°, (c) and (d) at 105°, (e) and (f) at 90°, (g) and (h) at 75°, and (i) and (j) at 60°.}
    \label{fig:fig8}
\end{figure*}

The velocity fields of the controlled flow for synthetic jets positioned at various angular orientations are shown in figure \ref{fig:fig8}, highlighting the sensitivity of control performance to jet placement.
Specifically, when the jets are positioned at 60°, the controlled flow still exhibits periodic vortex shedding, indicating that this orientation is less effective in stabilizing the flow. 
Conversely, when the jets are positioned at 120°, periodic vortex shedding is significantly suppressed, although some residual instabilities remain. 
This suggests that while the 120° orientation is more effective than 60°, it is not optimal. 
The most effective jet placements are at 75°, 90°, and 105°, where the wake of the cylinder flow is fully stabilized, and vortex shedding is completely eliminated.
These results underscore the critical role of jet orientation in achieving optimal flow control and highlight the potential for further optimization through precise placement of control actuators.

\subsubsection{Scenario A: Energy Consumption of Jet Pair }

\begin{figure*}
    \centering
    \includegraphics{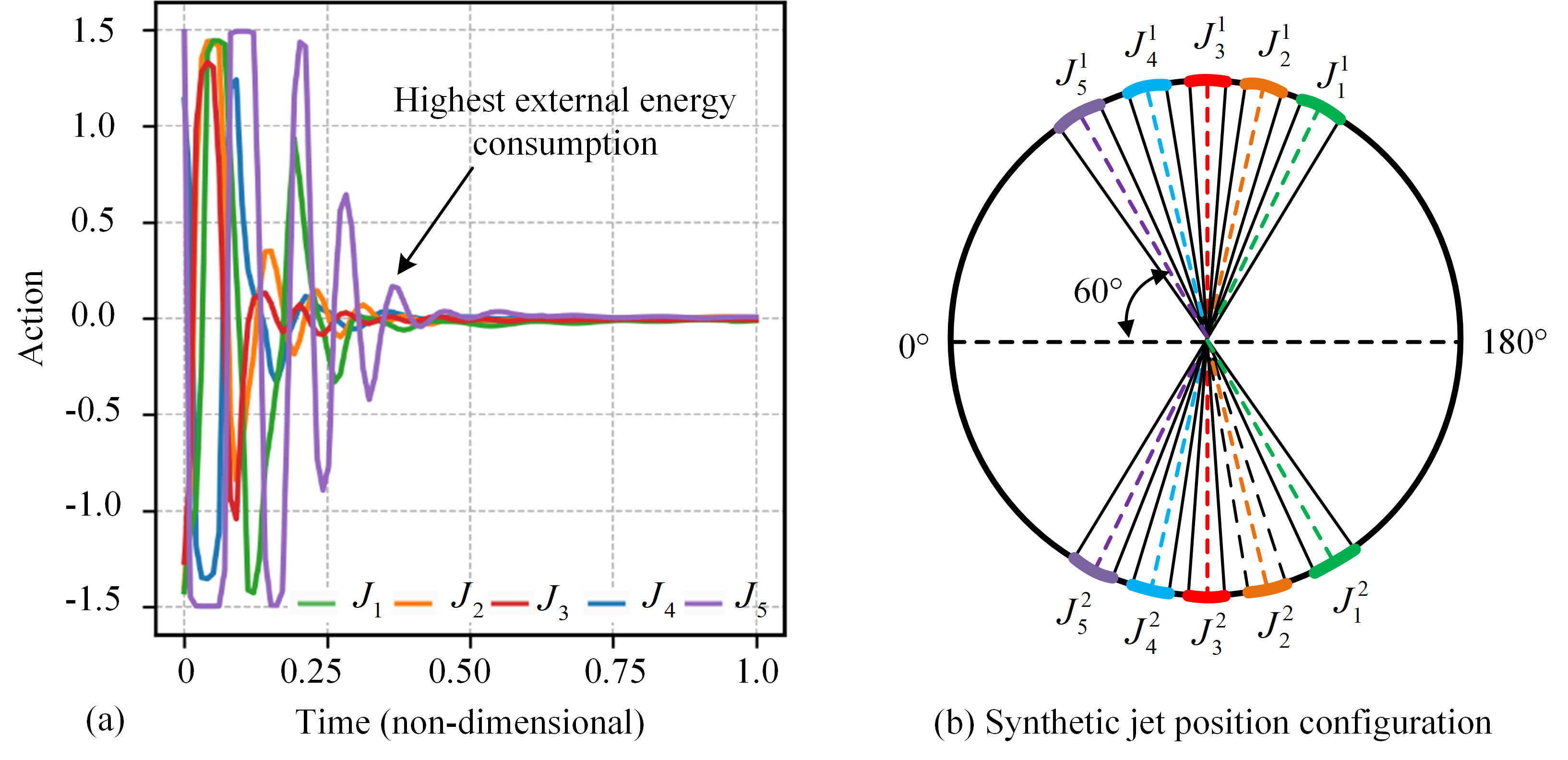}
    \caption{Energy consumption analysis of jet pairs positioned at different angular locations. (a) The control action values over non-dimensional time for each jet configuration. The integral of the control action values over time represents the total external energy consumed by the flow control system. (b) The the angular placement of jet pairs $J_1$ through $J_5$ around the cylinder.
    }
\label{fig:fig9}
\end{figure*}

The relationship between the control action values of each jet configuration and the dimensionless time is depicted in figure \ref{fig:fig9}, offering insights into the energy consumption required for the desired flow control. It also shows the azimuthal positions of five pairs of injectors along the surface of the cylinder. Among them, the jet position $ J_1 $ is located closest to the downstream stagnation point, while the jet position $ J_5 $ is situated closer to the upstream side. The control action value of the synthetic jet represents the intensity of the jet-driven force, and the integral of the control action value over time signifies the total external energy consumed to implement flow control.

As depicted in figure \ref{fig:fig9}, at the initial stage of flow control, the control effect curves of each jet configuration exhibit a peak, indicating the maximum energy required at the onset of jet actuation. Following this initial transient phase, the action values stabilize, suggesting that the control intensity and energy consumption become more stable and energy-efficient over time. Notably, the curves for $J_5$ and $J_1$ display the highest initial action values, indicating that these positions require more initial energy to initiate flow control compared to other configurations. In contrast, the action values for the synthetic jets positioned at $J_2$, $J_3$, and $J_4$ stabilize at a lower level than those of $J_5$ and $J_1$, demonstrating that these configurations are more energy-efficient in maintaining flow control.

In the later stages of flow control, the synthetic jet at $J_2$ operates effectively with only 1\% of the incoming flow rate, while those at $J_1$ and $J_5$ require over 3\% to achieve comparable control performance. This disparity suggests that injectors positioned at 60° and 120° exhibit lower energy efficiency, necessitating a higher input to sustain the desired control effect. In contrast, injectors at 90° and 105° demonstrate superior efficiency, maintaining effective flow control with significantly lower energy consumption. Such findings highlight the importance of selecting optimal jet positions to minimize energy expenditure while achieving targeted flow control objectives.
This insight highlights the critical importance of optimizing jet positioning to minimize energy consumption, a key consideration in designing efficient flow control systems that maximize aerodynamic performance while reducing power requirements around the cylinder.

\subsection{Multiple pairs of synthetic jets controllers}

\subsubsection{Scenario B: Control Performance of Multiple Jets Control}

Based on the results of flow control using a single pair of synthetic jets, it is observed that the configurations with jets positioned at 60° and 120°, namely $J_5$ and $J_1$, consume the most energy. Consequently, when employing multiple pairs of synthetic jets for flow control, the jets positioned at 60° and 120° are excluded. Instead, combinations involving $J_2$, $J_3$, and $J_4$ are explored. The tested configurations include two pairs of synthetic jets, specifically $J_2$ and $J_3$ as well as $J_3$ and $J_4$, and a configuration with three pairs, namely $J_2$, $J_3$, and $J_4$.

\begin{figure*}[htb!]
    \centering
    \includegraphics[width=\textwidth]{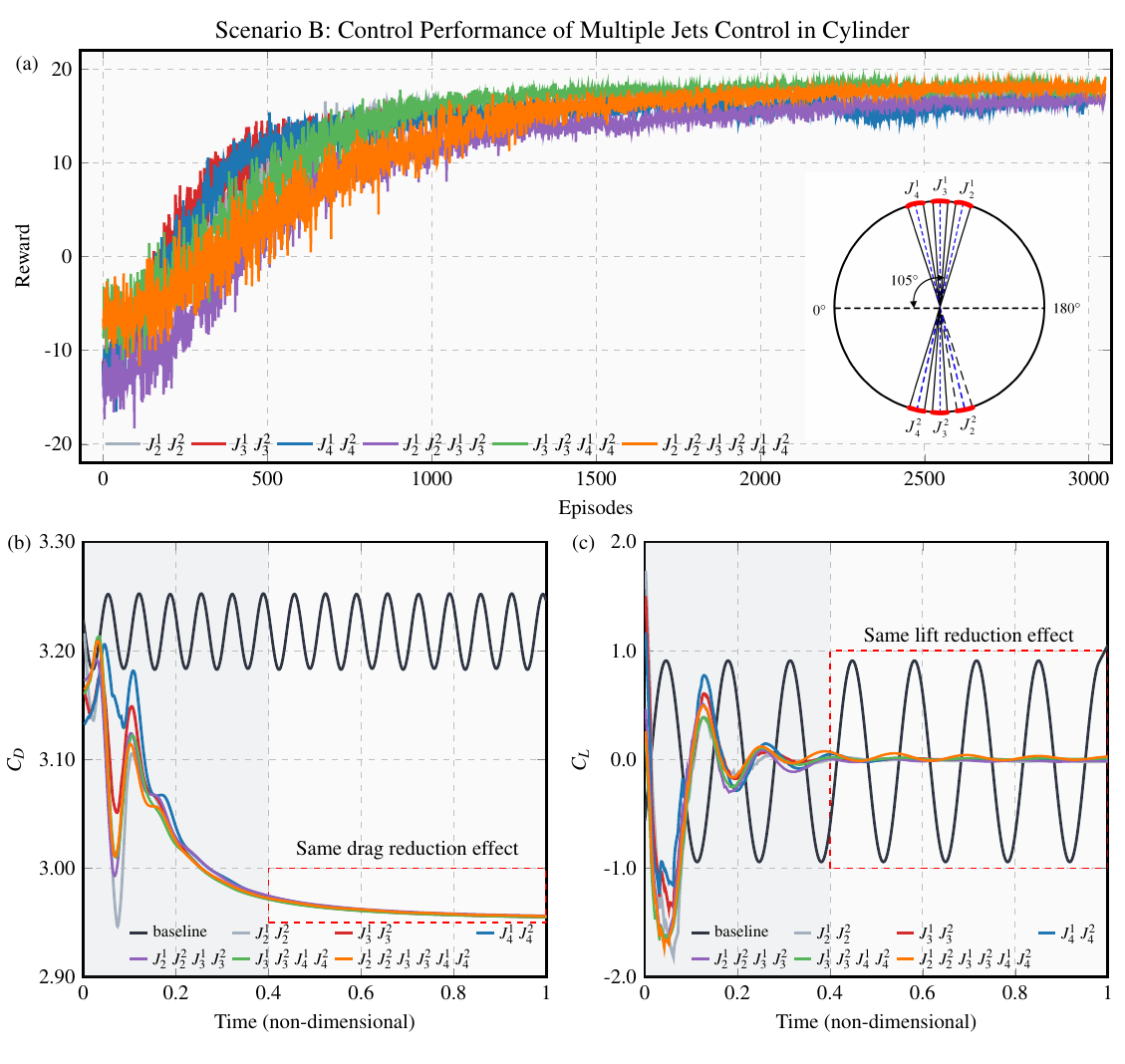}
    \caption{Control performance evaluation of multiple jet configurations around a cylinder.
    (a) Reward curves over training episodes for various multi-jet configurations. 
    (b) $C_D$ over non-dimensional time. 
    (c) $C_L$ over non-dimensional time.}
\label{fig:fig10}
\end{figure*}

In figure \ref{fig:fig10}, we present the control performance and convergence rates for flow control using multiple pairs of synthetic jets. To provide a comprehensive comparison, we also include the results of three single-pair synthetic jet controls, thereby contrasting the performance of multiple jet combinations with that of individual pairs. The reward curves in the figure reflect the effectiveness of the DRL agent in learning flow control strategies using different synthetic jet configurations. Specifically, figure \ref{fig:fig10}(a) offers a detailed analysis of the convergence performance during the training of deep reinforcement learning.
In all configurations, the reward function exhibits an increasing trend during the initial training phase. As the DRL training progresses, the reward values gradually stabilize, indicating that the agent has successfully optimized the control strategy. However, different jet configurations demonstrate varying convergence rates and maximum reward values.
Interestingly, we observe that employing multiple pairs of synthetic jets simultaneously for flow control does not accelerate the reinforcement learning training process. In fact, the reward functions for single-pair synthetic jets converge more rapidly than those for multiple pairs. This suggests that single-pair jet configurations are more conducive to faster convergence.

The variation of the drag coefficient with non-dimensional time for different jet configurations is illustrated in figure \ref{fig:fig10}(b). The curves indicate that all jet configurations achieve similar drag reduction, with the drag coefficient stabilizing after the initial transient phase. This stabilization suggests that the control strategies are effective in maintaining a consistent reduction in drag over time. 
The variation of the lift coefficient with non-dimensional time, demonstrating the effectiveness of the jet configurations in reducing lift fluctuations, is shown in figure \ref{fig:fig10}(c). 
The results reveal a significant reduction in lift oscillations across all jet configurations, with the lift coefficient stabilizing near zero after the initial transient phase. Similar to the drag coefficient, the consistent performance across configurations reinforces the conclusion that these jet positions are effective and efficient for flow control.
Additionally, we observe that the synthetic jets positioned at 105° ($J^1_1$ and $J^1_2$) exhibit a more pronounced transient response in both drag and lift coefficients, indicating a higher sensitivity to the jet actuation. This is a subtle but important observation that will be discussed further in subsequent sections.

\subsubsection{Scenario B: Energy Consumption with Simultaneous Control  (Two Pairs)}

\begin{figure*}
    \centering
    \includegraphics{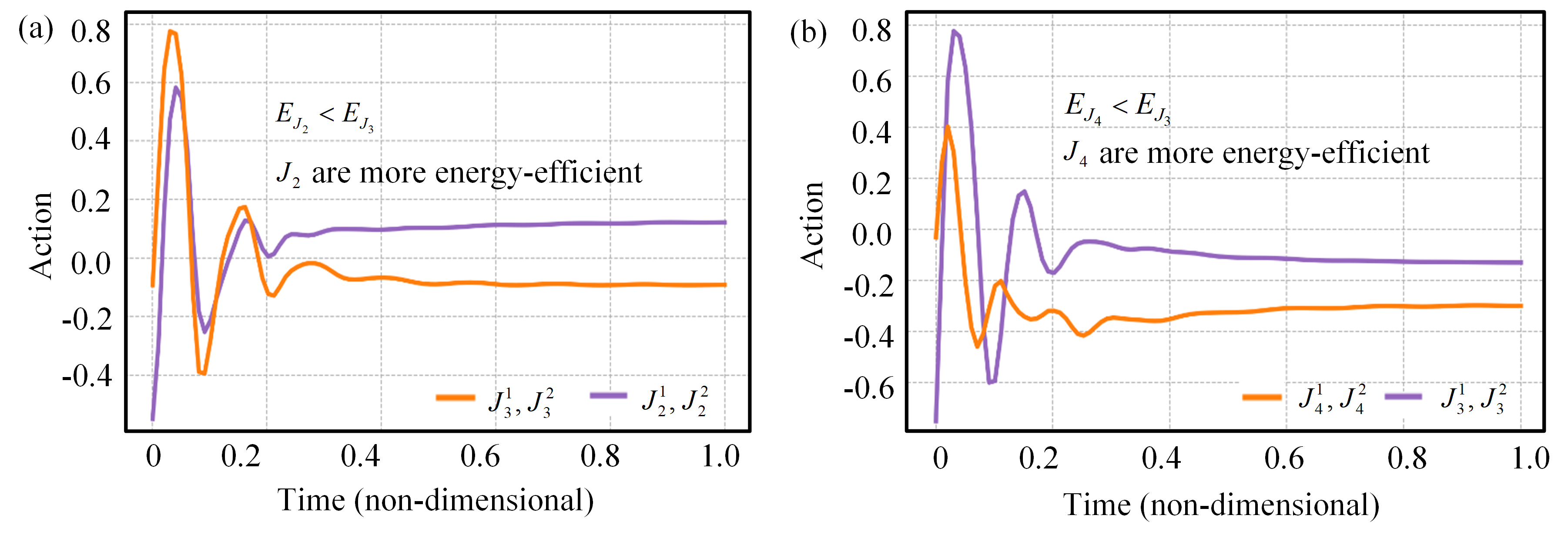}
    \caption{Comparison of action values over non-dimensional time for different synthetic jets configurations around the cylinder. (a) Two pairs of jets:  $J_2^1, J_2^2$ and $J_3^1, J_3^2$. (b) Two pairs of jets: $J_3^1, J_3^2$ and $J_4^1, J_4^2$.}
    \label{fig:fig11}
\end{figure*}

A flow control strategy involving the mass flow rates of two pairs of synthetic jets is implemented, which is referred to as “action” in figure \ref{fig:fig11}. The action values represent the external energy consumed to execute flow control, and the evolution of energy consumption for various jet configurations is plotted over dimensionless time.
Figure \ref{fig:fig11}(a) illustrates the energy consumption of the control strategy utilizing two pairs of synthetic jets: $J_3^1, J_3^2$ and $J_2^1, J_2^2$. The initial control actions display significant peaks, reflecting the high energy required at the onset of the flow control process. Over time, both jet configurations show a decrease in energy consumption, with the control efforts stabilizing.
However, the jet pair $J_3^1, J_3^2$ exhibits a slightly higher energy demand compared to the pair $J_2^1, J_2^2$. This indicates that the jets positioned at $J_2^1, J_2^2$ may be more energy-efficient in maintaining effective flow control.

Figure \ref{fig:fig11}(b) illustrates the control strategy using two pairs of synthetic jets, namely $J_3^1, J_3^2$ and $J_4^1, J_4^2$. The initial peak in the action values signifies the maximum energy required at the onset of the flow control process. Over time, the energy consumption stabilizes, indicating that the control strategy becomes more consistent. In the long term, the jet pair $J_4^1, J_4^2$ consumes slightly less energy compared to the jet pair $J_3^1, J_3^2$, suggesting that positioning the synthetic jets at $J_4$ may be more energy-efficient than jets at $J_3$.
In summary, the initial peak in the control strategy underscores the substantial energy demand at the inception of the process, while the subsequent stabilization reflects the sustained maintenance of flow control. Moreover, the disparities in energy consumption across different configurations provide valuable insights for optimizing flow control efficiency through judicious selection of jet placement. Specifically, jets positioned at $J_2$ locations exhibit superior energy efficiency compared to those at $J_3$ locations, and similarly, $J_4$ locations demonstrate enhanced energy savings relative to $J_3$ locations.

\subsubsection{Scenario B: Energy Consumption with Simultaneous Control  (Three Pairs)}

\begin{figure*}
    \centering
    \includegraphics{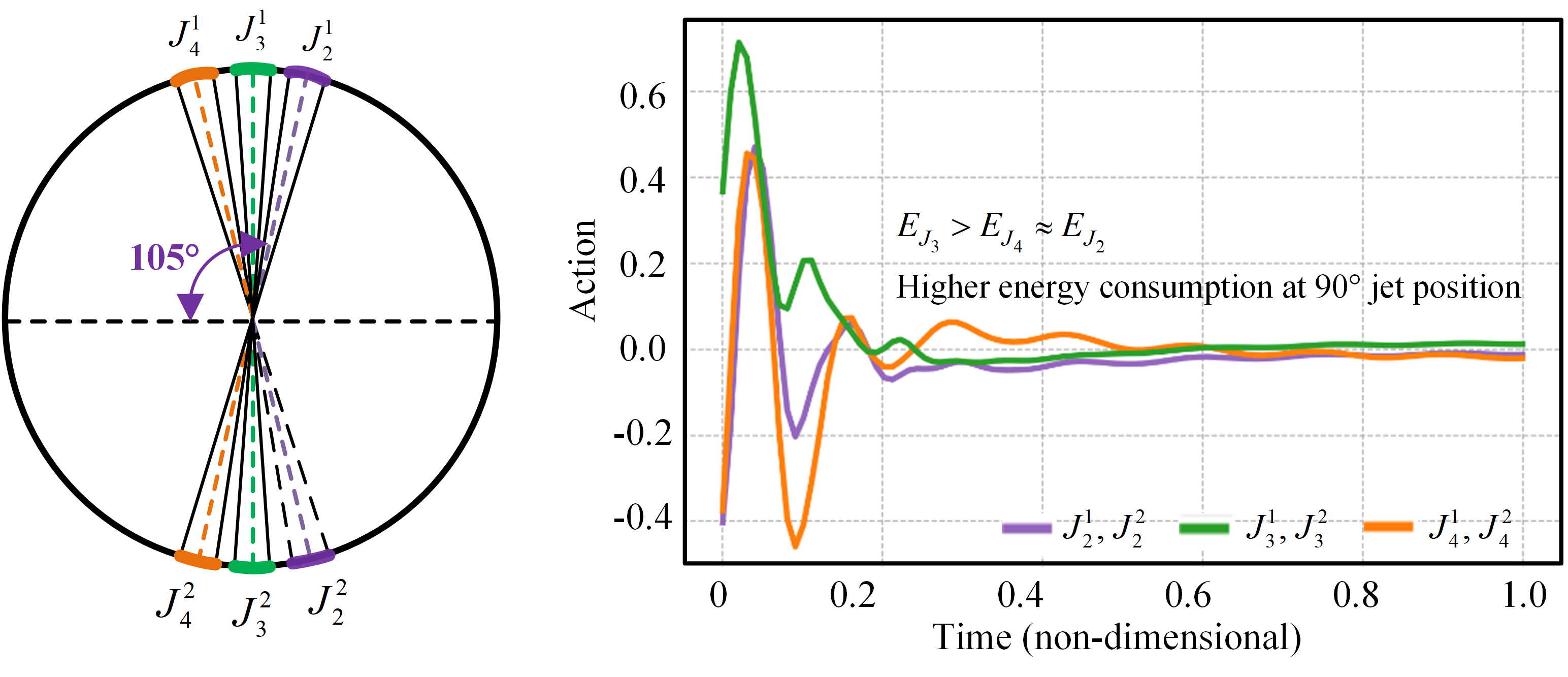}
    \caption{Energy consumption analysis for different jet pairs positioned around a cylinder. The left image shows the jet pairs $J_2$, $J_3$, and $J_4$, with azimuth angles ranging from 75° to 105°. The right image presents the corresponding actuation values of the synthetic jets over non-dimensional time for each jet configuration.}
    \label{fig:fig12}
\end{figure*}

Figure \ref{fig:fig12} presents a detailed analysis of energy consumption for the control strategy involving the simultaneous activation of three pairs of synthetic jets. The jets—\( J_2^1 \) and \( J_2^2 \), \( J_3^1 \) and \( J_3^2 \), and \( J_4^1 \) and \( J_4^2 \)—are symmetrically positioned around the cylinder at angular locations of 105°, 90°, and 75° from the forward stagnation point.
The external energy required for the synthetic jet pair \( J_3 \) (comprised of \( J^1_3 \) and \( J^2_3 \)) is denoted by \( E_{J_3} \), which is expressed as the integral of the control actions over time.
Interestingly, when compared to figure \ref{fig:fig11}, where only two pairs of synthetic jets are employed, figure \ref{fig:fig12} demonstrates that the activation of three pairs of synthetic jets leads to a reduction in the peak external energy demand for each individual jet. This suggests that the simultaneous use of additional jets can distribute the energy load more effectively, thereby enhancing the overall efficiency of the flow control strategy.

All three jet configurations exhibit pronounced peaks in the DRL-optimized control actions during the initial transient phase, followed by a stabilization of energy consumption over time. Among them, the configuration with jets located at $J_3$ shows the highest peak in control effort. The time-integrated control actions quantify the total energy input required for effective flow regulation, revealing that the energy consumption at $E_{J_3}$ is significantly higher than that of the other two configurations, with $E_{J_4} \approx E_{J_2}$.  
This observation suggests that the 90° jet placement demands greater energy input to achieve comparable control performance, indicating reduced energy efficiency relative to the $J_4$ and $J_2$ configurations. Moreover, between the latter two, jets at $J_2$ require less total energy than those at $J_4$, further highlighting $J_2$ as the most energy-efficient option. Figure~\ref{fig:fig12} summarizes the energy expenditure associated with different jet placements, emphasizing the importance of actuator positioning in achieving energy-efficient yet effective flow control.

Based on the analysis of DRL-based control strategies involving multiple pairs of synthetic jets, several key advantages over traditional flow control approaches are identified. One notable feature is that, regardless of whether two or three jet pairs are employed, no prior specification of actuation parameters such as frequency or amplitude is required. The control strategies are adaptively learned by the DRL agent through direct interaction with the flow field, eliminating the need for manual tuning. This highlights a fundamental distinction from conventional methods.
In comparison to traditional approaches, the DRL-based framework offers three major advantages. 
First, it removes the requirement for predefined control parameters, allowing the agent to autonomously infer effective strategies from flow feedback, thereby avoiding the exhaustive parameter studies typically required in periodic control schemes, such as those implemented by \citeauthor{Hongbo2024}. 
Second, it achieves superior control performance; while conventional techniques—e.g., those using control rods as studied by \citeauthor{LinLu2014}—may reduce drag or lift to some extent, they generally fail to fully suppress vortex shedding. In contrast, the DRL-based controller can stabilize the wake entirely. 
Third, unlike genetic algorithms such as that employed by \citeauthor{Zigunov2022}, which operate within predefined parameter spaces and risk missing global optima, DRL enables adaptive global exploration without prior knowledge of control laws. This results in significantly improved optimization efficiency and broader applicability to complex flow scenarios.

\section{Results: flow control performance of a square cylinder}\label{sec:sq}

In this section, we use the flow around a square cylinder as the DRL training environment, employing the control capabilities of a PPO agent to implement active flow control through synthetic jets positioned near the front and rear corners of the cylinder.
The flow around a square cylinder is characterized by earlier separation at sharp edges, more complex and turbulent wake structures, and greater flow instability compared to a circular cylinder, making it more challenging to control. This study aims to understand how the positioning of synthetic jets influences flow control performance and costs in this context. To explore these effects, we design four configurations with a single pair of synthetic jets, one configuration with two pairs of synthetic jets, and one additional configuration with a single pair of synthetic jets. These configurations are specifically designed to assess the impact of jet placement on the effectiveness and efficiency of flow control.

\subsection{A pair of synthetic jets controllers}

\subsubsection{Scenario A: Control Performance of Jet Pairs }

\begin{figure*}[htb!]
    \centering
    \includegraphics[width=\textwidth]{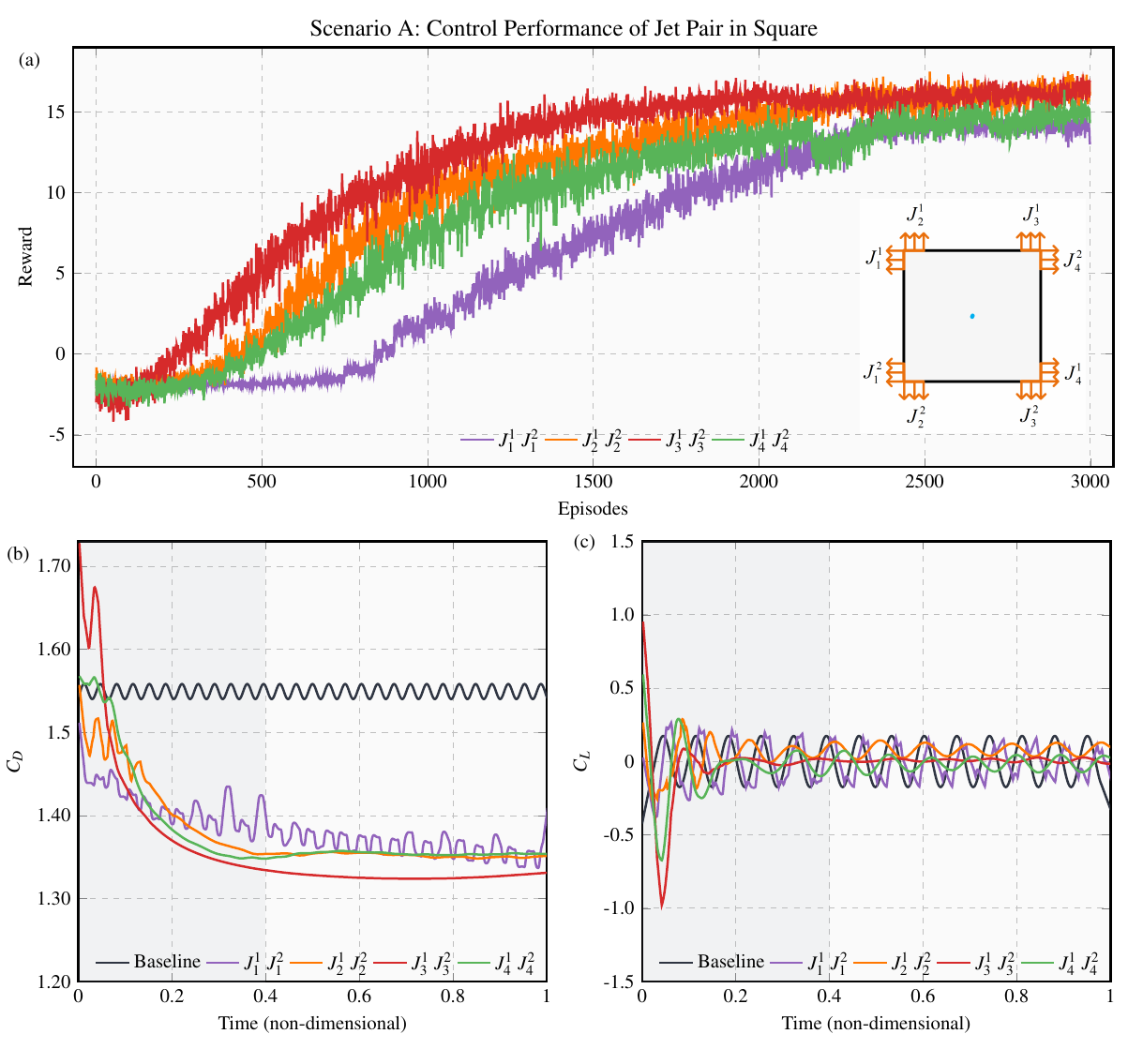}
    \caption{Performance evaluation of different jet configurations for flow control around a square cylinder. (a) reward curves over training episodes. (b) variation of $C_D$ over time, compared with the baseline scenario. (c) variation of $C_L$ over time, compared with the baseline scenario.}
    \label{fig:fig13}
\end{figure*}

DRL is used to develop flow control strategies for the flow around a square cylinder, as shown in figure \ref{fig:fig13}.
The figure provides a detailed analysis of the influence of synthetic jet positioning on flow control performance within the square geometry. The reward function serves as the primary metric for evaluating the effectiveness of the control strategies developed by the DRL agent. Initially, the reward values are low, reflecting the early stages of learning and the exploration of suboptimal strategies. As training progresses, a significant increase in the reward function is observed across all jet configurations, particularly between the 500 and 2000 episodes. This trend indicates that the DRL agent is successfully optimizing the control strategies, leading to improved flow control. The consistent rise in the reward function suggests that the DRL algorithm is steadily refining its policy, converging toward an optimal solution. 

The agent's ability to adapt and improve through continuous learning highlights the potential of DRL in addressing complex flow control challenges.
The reward function behavior during DRL training is examined with synthetic jets positioned near the front and rear corners of a square cylinder, specifically for the configurations $J_2^1, J_2^2$, $J_3^1, J_3^2$, and $J_4^1, J_4^2$, as illustrated in figure \ref{fig:fig13}(a). 
The results indicate that the reward function exhibited the most rapid growth when the jets are positioned at $J_3^1, J_3^2$, while the slowest growth was observed for the $J_2^1, J_2^2$ configuration.
This variation in growth rates underscores the superior adaptability of the $J_3^1, J_3^2$ configuration, as it allows the DRL agent to more effectively achieve the control objectives. The accelerated increase in the reward function associated with $J_3^1, J_3^2$ indicates that this setup is intrinsically more aligned with the flow control requirements, facilitating a more efficient optimization process.

\begin{table*}[htb!]
\centering
\caption{Control effectiveness and energy consumption analysis using a single pair of synthetic jets on a square cylinder.}
\begin{tabular*}{\textwidth}{>{\centering\arraybackslash}p{0.05\textwidth}
                            >{\centering\arraybackslash}p{0.06\textwidth}
                            >{\centering\arraybackslash}p{0.06\textwidth}
                            >{\centering\arraybackslash}p{0.12\textwidth}
                            >{\centering\arraybackslash}p{0.06\textwidth}
                            >{\centering\arraybackslash}p{0.06\textwidth}
                            >{\centering\arraybackslash}p{0.07\textwidth}
                            >{\centering\arraybackslash}p{0.07\textwidth}                            
                            >{\centering\arraybackslash}p{0.07\textwidth}
                            >{\centering\arraybackslash}p{0.12\textwidth}
                            }
\toprule
Jet & $C_{\text{D,Base}}$ & $C_{\text{D,Mean}}$ & Reduction (\%) & $C_{\text{D,Std}}$ & $C_{\text{L,Mean}}$ & $C_{\text{L,Std}}$ & $A_{\text{Std}}$ & $A_{\text{Mean}}$ & Action ratio (\%) \\
\hline
$J_1$ & 1.549 & 1.361 & 12.1 & 0.028 & 0.008 & 0.164 & 0.880  & 19.594 & 44.0 \\  
$J_2$ & 1.549 & 1.344 & 13.2 & 0.002 & 0.072 & 0.030 & -0.506 & 0.177 & 25.3 \\           
$J_3$ & 1.549 & 1.326 & 14.4 & 0.002 & 0.008 & 0.013 & -0.033 & 0.037 & 1.7 \\     
$J_4$ & 1.549 & 1.353 & 12.6 & 0.001 & -0.012 & 0.043 & 0.184 & 0.654 & 9.2 \\
\bottomrule
\end{tabular*}
\label{tab:table1}
\end{table*}

In contrast, the slower growth observed for $J_2^1, J_2^2$ suggests that this configuration may encounter challenges in interacting with the flow dynamics, potentially due to a more complex flow behavior that hinders the learning efficiency of the agent. These findings highlight the critical role of jet positioning in optimizing flow control strategies.
The distinct differences in reward growth rates not only reflect the varying degrees of learning efficiency but also emphasize the importance of strategic jet placement in enhancing the performance of DRL-based flow control. Configurations like $J_3^1, J_3^2$, which demonstrate rapid reward growth, are evidently more conducive to achieving efficient and precise flow manipulation, whereas slower growth configurations, such as $J_2^1, J_2^2$, may require more refined strategies or adjustments to overcome inherent complexities. 

The relationship between the drag coefficient and non-dimensional time is presented for four different jet placement configurations, compared to the baseline case without active control, as shown in figure \ref{fig:fig13}(b). The baseline curve, representing the drag coefficient without control, exhibits periodic oscillations. Once the synthetic jets are activated, the drag coefficient of the square cylinder decreases significantly across all configurations. The most notable reduction occurs when the jets are positioned near the rear corner points at $J_3^1, J_3^2$, indicating this placement is the most effective for drag reduction. In contrast, jets placed near the front corners at $J_1^1, J_1^2$ show a clear reduction in drag, but with transient spikes and oscillations, suggesting a dynamic interaction between the jets and the flow. Configurations at $J_2^1, J_2^2$ and $J_4^1, J_4^2$ also result in substantial drag reduction, though less pronounced than at $J_3^1, J_3^2$.

The relationship between the lift coefficient and non-dimensional time is depicted for various synthetic jet configurations, compared to the baseline scenario, as shown in figure \ref{fig:fig13}(c).
The baseline lift coefficient exhibits periodic fluctuations throughout the time series, characteristic of the natural vortex shedding around the square cylinder.
When the synthetic jets are positioned at $J_2^1, J_2^2$, $J_3^1, J_3^2$, and $J_4^1, J_4^2$, a significant attenuation in the lift coefficient is observed. Specifically, the configuration at $J_3^1, J_3^2$ induces the most pronounced initial reduction in the lift coefficient, which rapidly stabilizes near zero, indicating effective suppression of the unsteady aerodynamic forces.
Furthermore, the configurations at $J_2^1, J_2^2$ and $J_4^1, J_4^2$ lead to a marked reduction in both the amplitude and frequency of the lift coefficient oscillations, thereby significantly damping the lift fluctuations. This suggests a successful mitigation of vortex-induced forces through controlled jet actuation.
However, when the jets are located at $J_1^1, J_1^2$, the lift coefficient fails to exhibit a similar reduction. Instead, transient spikes and persistent oscillations are present, indicating that this jet configuration is less effective in altering the flow structures responsible for lift generation.
These results crystallize the crucial role of optimal jet placement in flow control, where precise modulation of aerodynamic forces, such as lift, is essential for achieving effective performance.

\subsubsection{Scenario A: Energy Consumption of Jet Pairs }

\begin{figure*}[htb!]
    \centering
    \includegraphics[width=0.95\textwidth]{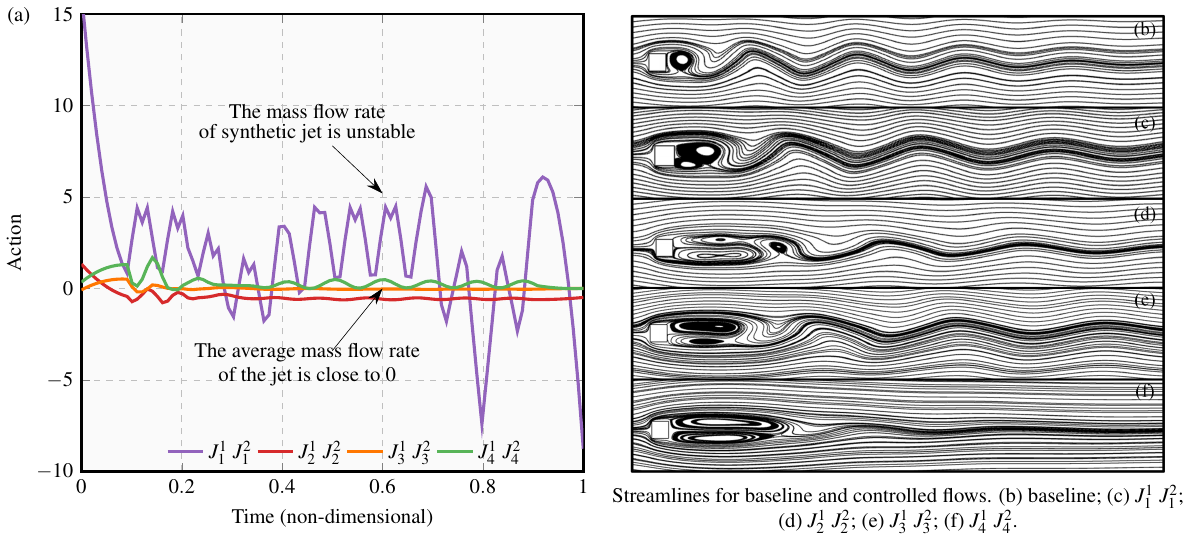}
    \caption{(a) Energy consumption over non-dimensional time for jet pairs positioned at various locations around the square cylinder. (b) Streamlines of the controlled flow and the baseline flow when the synthetic jet is located at different positions.}
    \label{fig:fig14}
\end{figure*}

Figure \ref{fig:fig14} presents a comparative visualization of the synthetic jet actuation history and corresponding flow structures for selected jet configurations. 
Figure \ref{fig:fig14} (a) illustrates the temporal evolution of the control action, represented as the normalized mass flow rate of synthetic jets, for four distinct jet placements denoted as $J_1^1 J_1^2$, $J_2^1 J_2^2$, $J_3^1 J_3^2$, and $J_4^1 J_4^2$.
The control input, expressed as the non-dimensionalized mass flow rate of the synthetic jet, is actively modulated by the reinforcement learning agent.
All control signals exhibit oscillatory behavior with zero mean, a typical feature of synthetic jets. However, the amplitude, frequency, and structure of the control inputs vary significantly depending on the jet placement.

The external energy consumed for flow control varies significantly with the locations of the synthetic jets at $ J_1^1, J_1^2 $, $ J_2^1, J_2^2 $, $ J_3^1, J_3^2 $, and $ J_4^1, J_4^2 $. When the jets are positioned at $ J_1^1 $ and $ J_1^2 $, the energy required for flow control increases markedly, accompanied by pronounced fluctuations, indicating unstable mass flow rates. This instability suggests that the complex interactions with the flow field result in lower control efficiency, making consistent and effective flow manipulation challenging.
In contrast, when the jets are located at the other positions, the control strategy exhibits more stable behavior with minimal oscillations and values close to zero, reflecting efficient and balanced flow control with lower energy input. Specifically, the fluctuation amplitudes of the control actions at $ J_2^1, J_2^2 $ and $ J_4^1, J_4^2 $ are notably smaller than those at $ J_1^1, J_1^2 $.
Notably, the configuration with jets at $ J_3^1 $ and $ J_3^2 $ requires the least energy at the onset of control and demonstrates the most stable control actions over time. The comparison between the less stable $ J_1^1, J_1^2 $ configuration and the more stable $ J_3^1, J_3^2 $ configuration highlights the differences in energy efficiency and control stability.

The differences in the observed control signals correspond to distinct flow modifications illustrated in figure \ref{fig:fig14}(a), where the streamlines reveal varying degrees of vortex suppression.  
The configuration with the highest energy consumption, $J_1^1 J_1^2$, results in the least effective flow control, as shown in figure \ref{fig:fig14}(c).  
Moderate energy-consuming placements, $J_2^1 J_2^2$ and $J_4^1 J_4^2$, yield improved wake stability compared to the baseline case, as depicted in figures \ref{fig:fig14}(d) and \ref{fig:fig14}(e), respectively.  
The most energy-efficient configuration, $J_3^1 J_3^2$, leads to the most stabilized wake structure, as shown in figure \ref{fig:fig14}(f), where the wake length is significantly shortened and the vortex intensity is markedly reduced, demonstrating superior control performance.  
In contrast, other jet placements induce less pronounced modifications in the wake.  
These findings highlight the critical importance of jet placement in determining the effectiveness of synthetic jet-based flow control strategies.

\begin{figure*}[htb!]
    \centering
    \includegraphics{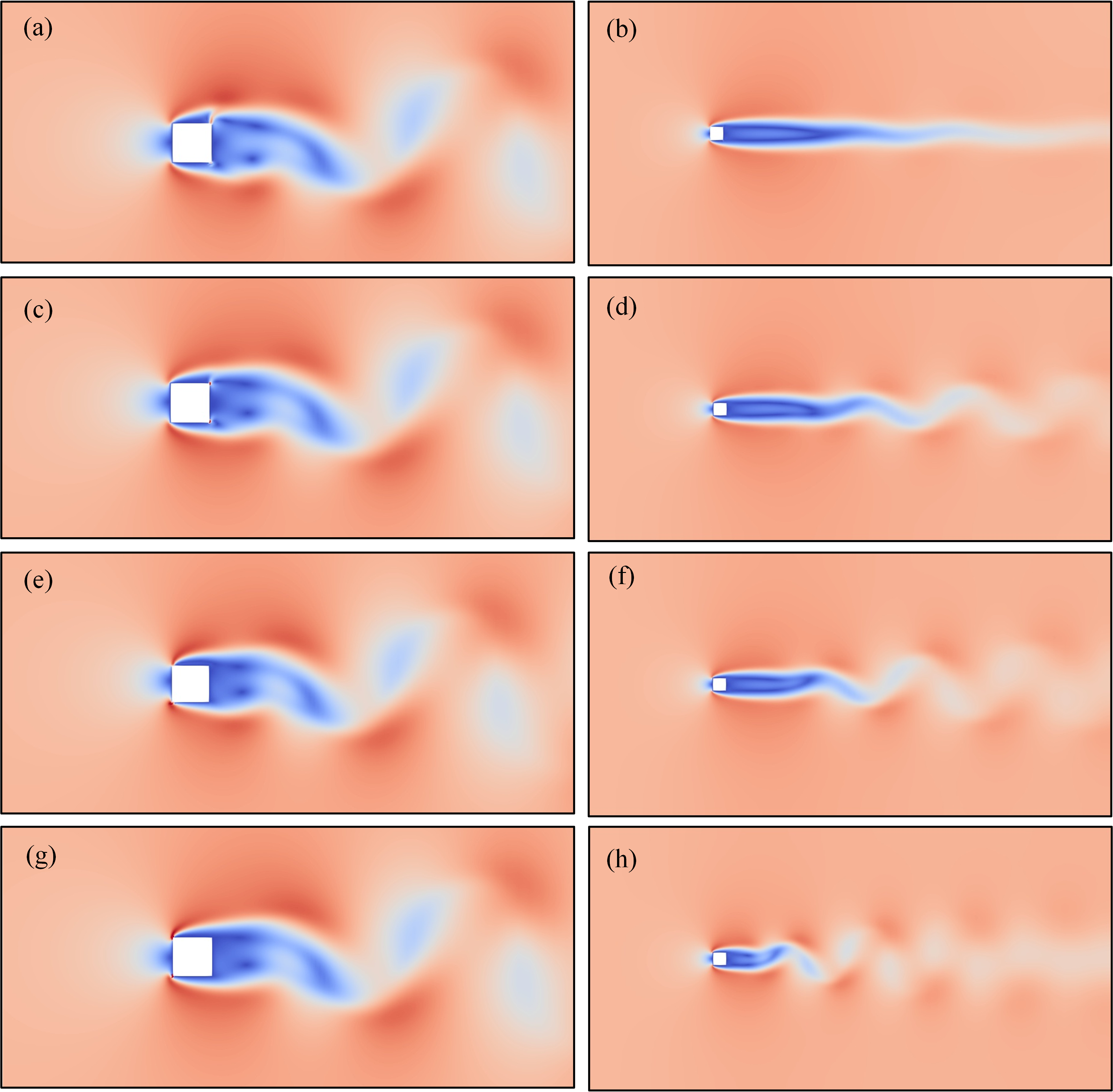}
    \caption{ Velocity contours of flow around a square cylinder with synthetic jets positioned at various locations. The left column shows the flow with active jet control, while the right column presents the corresponding flow control results.
    (a) and (b) Jets $J_3^1$ and $J_3^2$ on the top and bottom sides near the rear corners.
    (c) and (d) Jets $J_4^1$ and $J_4^2$ on the rear side.
    (e) and (f) Jets $J_2^1$ and $J_2^2$ on the top and bottom sides near the front corners.
    (g) and (h) Jets $J_1^1$ and $J_1^2$ on the front side.}
    \label{fig:fig15}
\end{figure*}

Figure \ref{fig:fig15} presents the activation states of the synthetic jets at positions $J_1$, $J_2$, $J_3$, and $J_4$, alongside the corresponding velocity contours of the controlled flow field. The left column illustrates the instantaneous velocity fields with active jet actuation, while the right column depicts the resulting wake structures.  
The velocity contours reveal that when the synthetic jets are positioned near the rear corners of the square cylinder at $J_3$, vortex shedding is entirely suppressed, yielding a highly stabilized wake.
The synthetic jets at $J_4$, located on the rear face of the cylinder, the recirculation region expands significantly compared to the initial stages of control. 
The synthetic jets at $J_2$, where the jets are located near the front corners, periodic actuation interacts with the shear layers, mitigating flow separation and enhancing wake. 
Finally, when the jets are positioned at $J_1$ on the front face, their interaction with the incoming free stream improves near-wake stability; however, alternating vortex shedding persists, suggesting that upstream jet placement is less effective in fully eliminating wake instabilities.  
These results underscore the critical role of jet positioning in flow control efficacy, demonstrating that optimal placement at $J_3$ achieves complete vortex suppression, whereas other configurations yield varying degrees of wake stabilization and turbulence modulation.

The control performance and energy consumption of flow control based on DRL are summarized in table \ref{tab:table1}. The control performance is evaluated by the percentage reduction in drag coefficient $ C_{\text{D,Mean}} $ relative to the baseline drag coefficient $ C_{\text{D,Base}} $, the standard deviation of the drag coefficient $ C_{\text{D,Std}} $ for the controlled flow, and the mean $ C_{\text{L,Mean}} $ and standard deviation $ C_{\text{L,Std}} $ of the lift coefficient.
The energy consumption for flow control is quantified by the mean $ A_{\text{Mean}} $ and standard deviation $ A_{\text{Std}} $ of the jet actuation magnitude, as well as the action ratio, which is the percentage of the average jet flow rate relative to the upstream inlet flow rate.
The quantified control performance and energy consumption metrics in the table indicate that positioning the synthetic jet at $ J_3 $ achieves the highest drag reduction efficiency of 14.4\%, using only 1.7\% of the inlet flow rate, while also minimizing the standard deviations of $ C_D $, $ C_L $, and jet actuation.
In contrast, placing the jet at $ J_1 $ requires 44\% of the inlet flow rate to achieve a drag reduction of only 12.1\%. This configuration also exhibits the highest standard deviations in $ C_D $, $ C_L $, and jet actuation, indicating the least stability.

Overall, the control performance of the jets varies significantly with placement. When the synthetic jets are positioned near the front corner points, vortex shedding is not fully suppressed due to the large separated flow reattaching along the sides of the square cylinder before separating again near the rear corners. In contrast, placing the jets near the rear corner points at position $J_3$ provides more effective flow control, leading to the complete suppression of vortex shedding.
\citeauthor{chen2023deep} used DRL-based flow control to reduce vortex-induced vibrations around a square cylinder and compared the control effects of jet placement at the front, middle, and rear positions along the cylinder's side. Their results showed that full suppression was achieved only near the rear corners, consistent with our findings \cite{chen2023deep}.

\subsection{Multiple pairs of synthetic jets controllers}

\subsubsection{Scenario B: Control Performance of Multiple Jets Control}

\begin{figure*}[htb!]
    \centering
    \includegraphics[width=\textwidth]{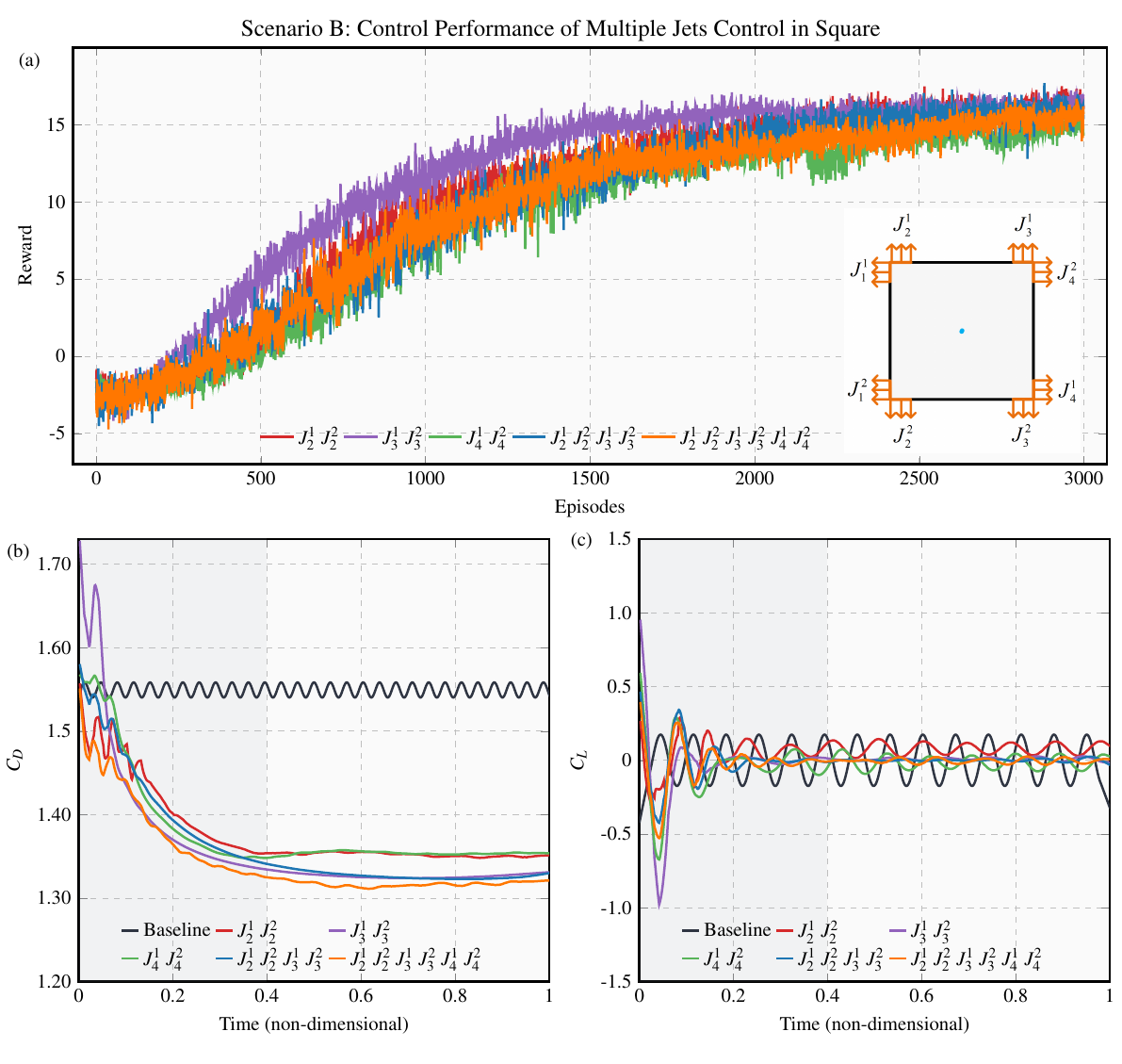}
    \caption{Evaluation of flow control performance for a square cylinder using single and multiple synthetic jets configurations. (a) Reward curves during the training process. (b) $ C_D $ over non-dimensional time. (c) $ C_L $ over non-dimensional time. }
    \label{fig:fig16}
\end{figure*}

To further explore the influence of actuator placement on flow control performance, additional experiments were conducted by simultaneously employing multiple pairs of synthetic jets. The corresponding results, including the training process of the DRL agent and its flow control effectiveness for a square cylinder with various jet configurations, are presented in figure~\ref{fig:fig16}. For comparison, the performance achieved using a single pair of synthetic jets is also included. Figure~\ref{fig:fig16}(a) illustrates the temporal evolution of the reward function throughout the training process for different jet configurations. In particular, the configurations $ J^1_2 J^2_2 $, $ J^1_3 J^2_3 $, and $ J^1_4 J^2_4 $ denote specific synthetic jet locations on the square cylinder surface, and their comparative performance underscores the impact of jet placement on learning efficiency and flow control capability.

Across all four configurations, the reward functions increase as training progresses, indicating that the DRL agent is successfully learning and refining the flow control strategies. Initially, the reward values are low, reflecting the early stages of learning when the agent explores various control strategies. As the number of episodes increases, the reward function for each jet configuration shows steady improvement, particularly between 500 and 2000 episodes. The gradual rise in the reward values suggests that the DRL algorithm is progressively enhancing its strategy, leading to improved control performance.
Crucially, the reward function for the $J^1_3 J^2_3$ configuration exhibits faster growth, indicating that this setup is particularly effective in achieving the control objectives. The quicker convergence suggests that the agent is learning the optimal control strategy more efficiently with this configuration. Conversely, the more gradual increase in reward functions observed when using multiple pairs of synthetic jets suggests a more incremental learning process. The slower convergence may result from increased complexity in the fluid dynamics or suboptimal jet placement for these specific control tasks.

As a function of non-dimensional time for various jet configurations, compared to a baseline scenario without active flow control, the drag coefficient is illustrated in figure \ref{fig:fig16}(b).
The baseline drag coefficient exhibits periodic oscillations, a typical characteristic of unsteady flow around a square cylinder without control.
Upon activating the synthetic jets, the drag coefficient for all configurations decreases significantly, indicating the effectiveness of the control strategies. 
Specifically, the drag reduction achieved by single jet pairs positioned at $J_2^1, J_2^2$ and $J_4^1, J_4^2$ is less pronounced than that of the jet pair at $J_3^1, J_3^2$. When both $J_2^1, J_2^2$ and $J_3^1, J_3^2$ are activated simultaneously, the drag reduction is almost identical to that achieved by $J_3^1, J_3^2$ alone, suggesting that adding the second jet pair ($J_2^1, J_2^2$) does not provide additional benefit.
However, when all three jet pairs ($J_2^1, J_2^2$, $J_3^1, J_3^2$, and $J_4^1, J_4^2$) are activated together, there is a more substantial reduction in drag compared to both the single jet pair ($J_3^1, J_3^2$) and the two jet pairs ($J_2^1, J_2^2$ and $J_3^1, J_3^2$). This enhanced drag reduction comes at the cost of some stability, reflecting the more complex interactions involved in controlling the flow with multiple jet pairs.
This analysis highlights the trade-offs between achieving greater drag reduction and maintaining stability when using multiple synthetic jets in flow control. The results suggest that while adding more jets can further reduce drag, it may also introduce additional challenges in maintaining a stable control strategy.

Over non-dimensional time for various jet configurations, compared to a baseline scenario without active control, the lift coefficient is illustrated in figure \ref{fig:fig16}(c).
The baseline $C_L$ without control exhibits periodic oscillations. 
The activation of synthetic jets significantly reduces both the amplitude and frequency of these oscillations, stabilizing the lift coefficient closer to zero or the desired level. This suppression of lift fluctuations helps create a more stable flow field by mitigating the unsteady forces acting on the cylinder. 
Specifically, the configuration with jets positioned at $J_3^1, J_3^2$ shows a more pronounced reduction in lift compared to those at $J_2^1, J_2^2$ or $J_4^1, J_4^2$. When two jet pairs ($J_2^1, J_2^2$ and $J_3^1, J_3^2$) are activated simultaneously, the reduction in lift is similar to that achieved with $J_3^1, J_3^2$ alone, indicating that the additional jet pair does not significantly enhance control.

When all three pairs of synthetic jets ($J_2^1, J_2^2$, $J_3^1, J_3^2$, and $J_4^1, J_4^2$) are activated simultaneously, the lift coefficient exhibits intensified oscillations, signifying increased flow instability. Although this configuration leads to notable drag reduction, the resulting unsteady aerodynamic forces may undermine the overall robustness of the control strategy. This highlights a fundamental trade-off between achieving lower drag and maintaining stable lift forces.  
A comparative analysis of different jet configurations further elucidates this trade-off. Among the tested configurations, the $J_3^1, J_3^2$ placement proves to be the most effective, demonstrating rapid reward function convergence, significant drag reduction, and stabilized lift dynamics. In contrast, while the simultaneous activation of two jet pairs ($J_2^1, J_2^2$ and $J_3^1, J_3^2$) yields a comparable level of drag mitigation, it fails to match the convergence efficiency achieved by the $J_3^1, J_3^2$ configuration alone.
Furthermore, while the deployment of all three jet pairs amplifies drag reduction, the induced lift fluctuations underscore the trade-off between aerodynamic performance and flow stability. This interplay between control effectiveness and induced instabilities will be further examined in the subsequent section, with a specific focus on the energy expenditure associated with different jet actuation strategies.

\subsubsection{Scenario B: Energy Consumption with Simultaneous Control  (Two Pairs and Three Pairs)}

\begin{figure*}[htb!]
    \centering
    \includegraphics{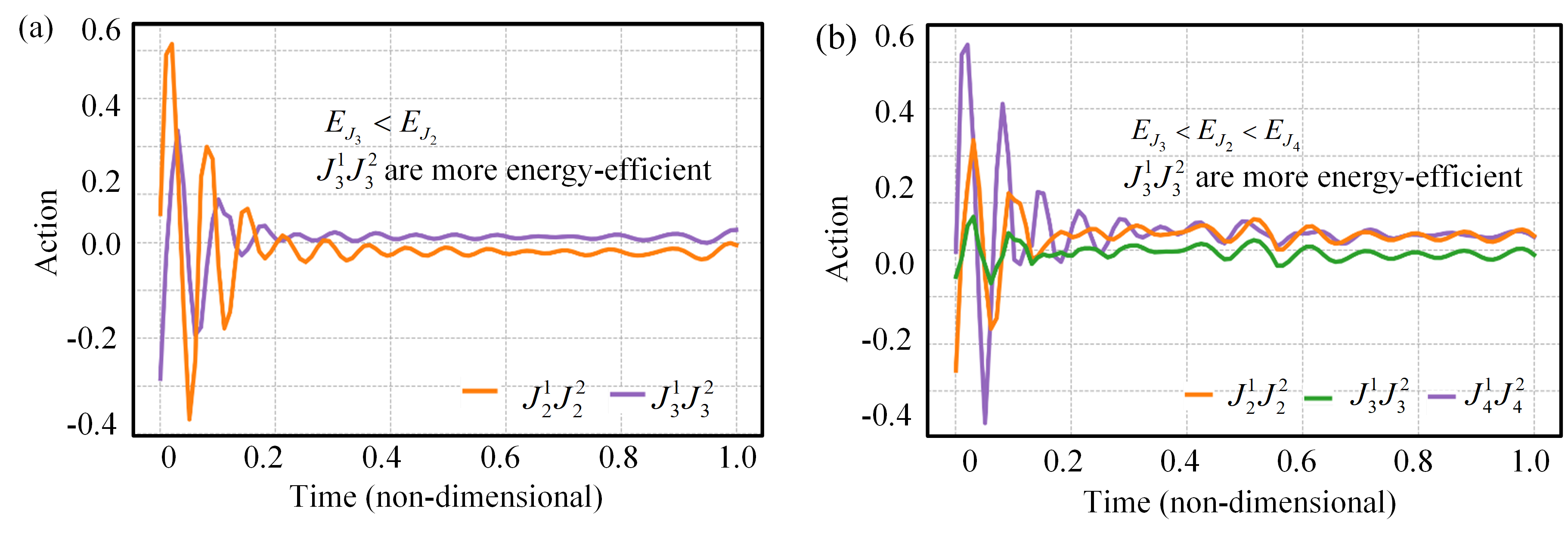}
    \caption{Comparison of energy consumption during flow control using different synthetic jets configurations around a square cylinder. (a) Energy consumption with two jet pairs $J_2^1, J_2^2$ and $J_3^1, J_3^2$. (b) Energy consumption with three jet pairs $J_2^1, J_2^2$, $J_3^1, J_3^2$, and $J_4^1, J_4^2$.}
    \label{fig:fig17}
\end{figure*}

The external energy consumption over non-dimensional time for flow control with two (a) and three (b) synthetic jets pairs around a square cylinder is depicted in figure \ref{fig:fig17}.
When comparing the single-pair synthetic jets control in figure \ref{fig:fig14} to the multi-pair configuration in figure \ref{fig:fig17}, it is evident that the energy consumption in the multi-jet scenario is reduced by an order of magnitude relative to the single-jet case. 
Where two pairs of synthetic jets, $J_2^1, J_2^2$ and $J_3^1, J_3^2$, are simultaneously employed, significant peaks in action values for both configurations are shown in the initial phase in figure \ref{fig:fig17}(a), indicating a high energy demand at the onset of flow control.
This peak reflects the system response to the abrupt introduction of control forces generated by the jets.
As time progresses, the action values stabilize, signifying that the energy required to maintain control has reached a steady state. 

The total energy consumption of the synthetic jets located at $J_3^1$ and $J_3^2$ is significantly lower than that of the jets positioned at $J_2^1$ and $J_2^2$, highlighting the superior energy efficiency of the $J_3$ configuration.  
This indicates that the $J_3^1$ and $J_3^2$ jet configuration achieves the desired flow control with reduced external energy input, making it a more effective option for long-term operation.

Figure \ref{fig:fig17}(b) presents the control performance when three pairs of synthetic jets (\(J_2^1, J_2^2\); \(J_3^1, J_3^2\); and \(J_4^1, J_4^2\)) are simultaneously deployed around the square cylinder. The control action quantified by the time integral of actuation signals, reflects the total energy expenditure required to implement flow control via synthetic jets. A lower cumulative actuation input corresponds to reduced external energy consumption, thereby indicating enhanced energy efficiency. As in the two-jet scenarios, all jets exhibit elevated actuation levels during the initial transient phase, reflecting the higher energy demand required to rapidly suppress vortex shedding and stabilize the wake. Once the control effect is established, the actuation levels approach a quasi-steady state.

Among the three configurations, the jets located at \(J_4\) maintain the highest sustained action values, indicating a greater continuous energy requirement for maintaining effective control. In contrast, the jets at \(J_3\) consistently exhibit lower energy expenditure throughout the control horizon. This observation not only corroborates their superior performance in the two-jet configuration but also highlights their robustness in more complex control setups. The persistent energy efficiency of the \(J_3\) configuration across both two- and three-jet arrangements suggests that this location—likely corresponding to critical regions such as separation or reattachment points—offers an optimal balance between control authority and energy cost.
These findings are consistent with prior results obtained from single-jet analysis, where jets at \(J_3\) demonstrated effective suppression of unsteady wake dynamics with minimal actuation. Overall, the analysis emphasizes the critical role of jet placement in achieving energy-efficient flow control, with \(J_3\) emerging as the most favorable location for long-term, low-energy operation.

\section{Discussion: the physical insight of flow control strategy}\label{sec:Discussion} 

\subsection{Flow separation phenomenon}

\begin{figure*}[htb!]
    \centering
    \includegraphics{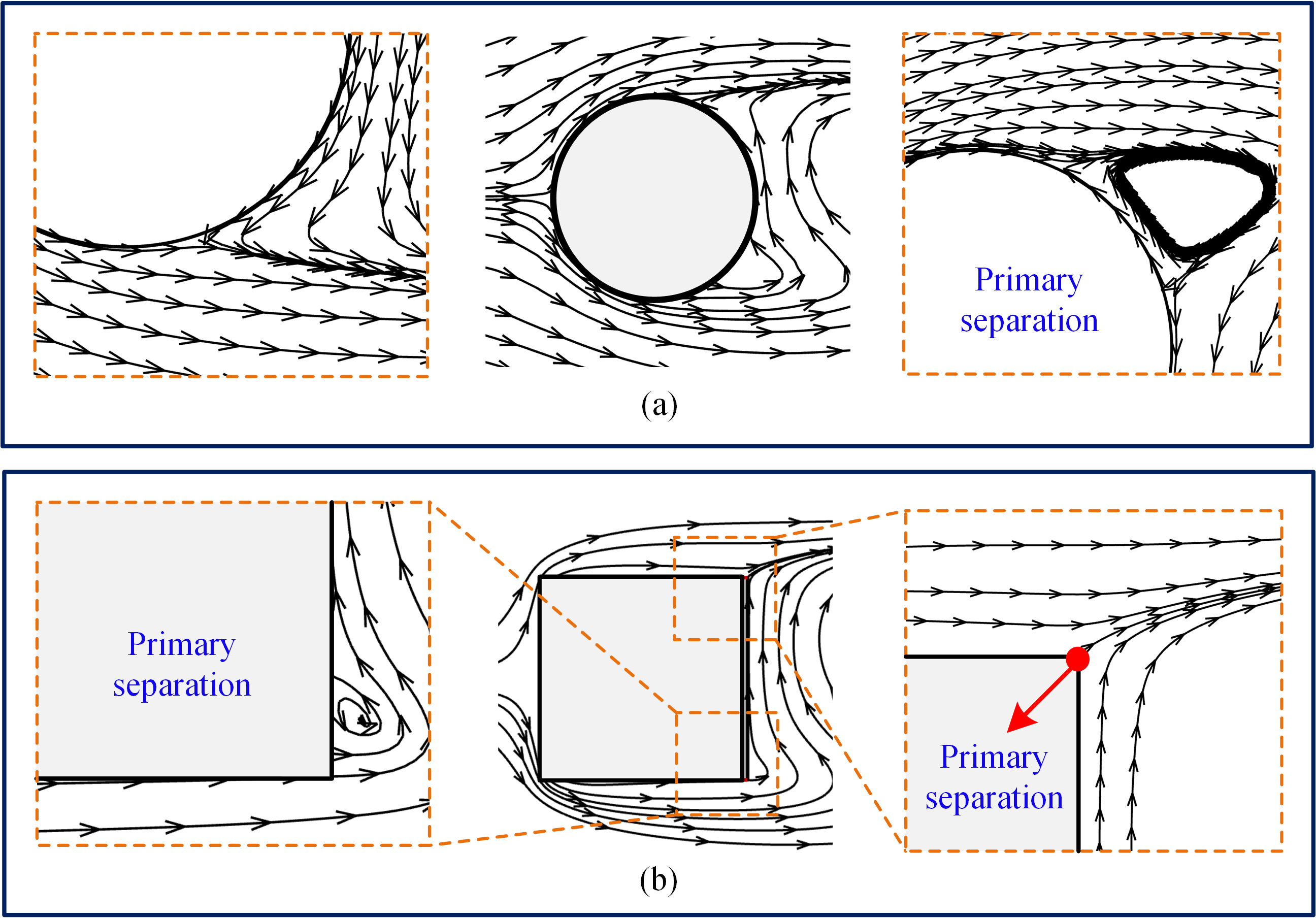}
    \caption{Streamline pattern for the baseline flow at $ Re = 100 $. (a) circular cylinder and (b) square cylinder.}
    \label{fig:fig18}
\end{figure*}

Section \ref{sec:cy} and section \ref{sec:sq} provide a detailed examination of the effects of various jet configurations on flow control performance. 
To further analyze the underlying mechanisms of flow control in relation to fluid dynamics, we present a comprehensive discussion of the control performance for both circular and square cylinder flows.
When $ Re = 100 $, the flow around the circular cylinder exhibits a stable laminar characteristic with well-defined separation points, as shown in figure \ref{fig:fig18}. Flow separation occurs symmetrically on both sides of the cylinder, approximately 105° from the forward stagnation point. At this location, the adverse pressure gradient causes the boundary layer to detach from the surface, marking the onset of flow separation. This detachment initiates the formation of a wake behind the cylinder, where alternating vortices periodically shed, ultimately leading to the development of the characteristic von Kármán vortex street downstream.

In contrast, for a square cylinder at the same Reynolds number, the sharp changes in geometry result in flow separation occurring at the four sharp corners of the body. The boundary layer detaches immediately at these edges, forming distinct vortices. Especially, large separation bubbles form near the rear corners, creating a pronounced separation zone in the wake.
The differing nature of flow separation around circular and square cylinders necessitates distinct control strategies, posing unique challenges when applying reinforcement learning algorithms for flow control.

\subsection{Flow control mechanism}

Through comparative analysis, the optimal position for synthetic jets around a circular cylinder is found to be at a 105° angle from the front stagnation point. At this position, drag and lift are significantly reduced, vortex shedding is fully suppressed, and energy consumption is minimized. This result aligns with the findings of \cite{wang2024dynamic} at $Re = 100$, which demonstrated superior control performance using a sensor at 105°. The results further confirm that the 105° configuration offers the best balance between energy efficiency and flow stability, as detailed in section \ref{sec:cy}.

\begin{figure*}[htb!]
    \centering
    \includegraphics{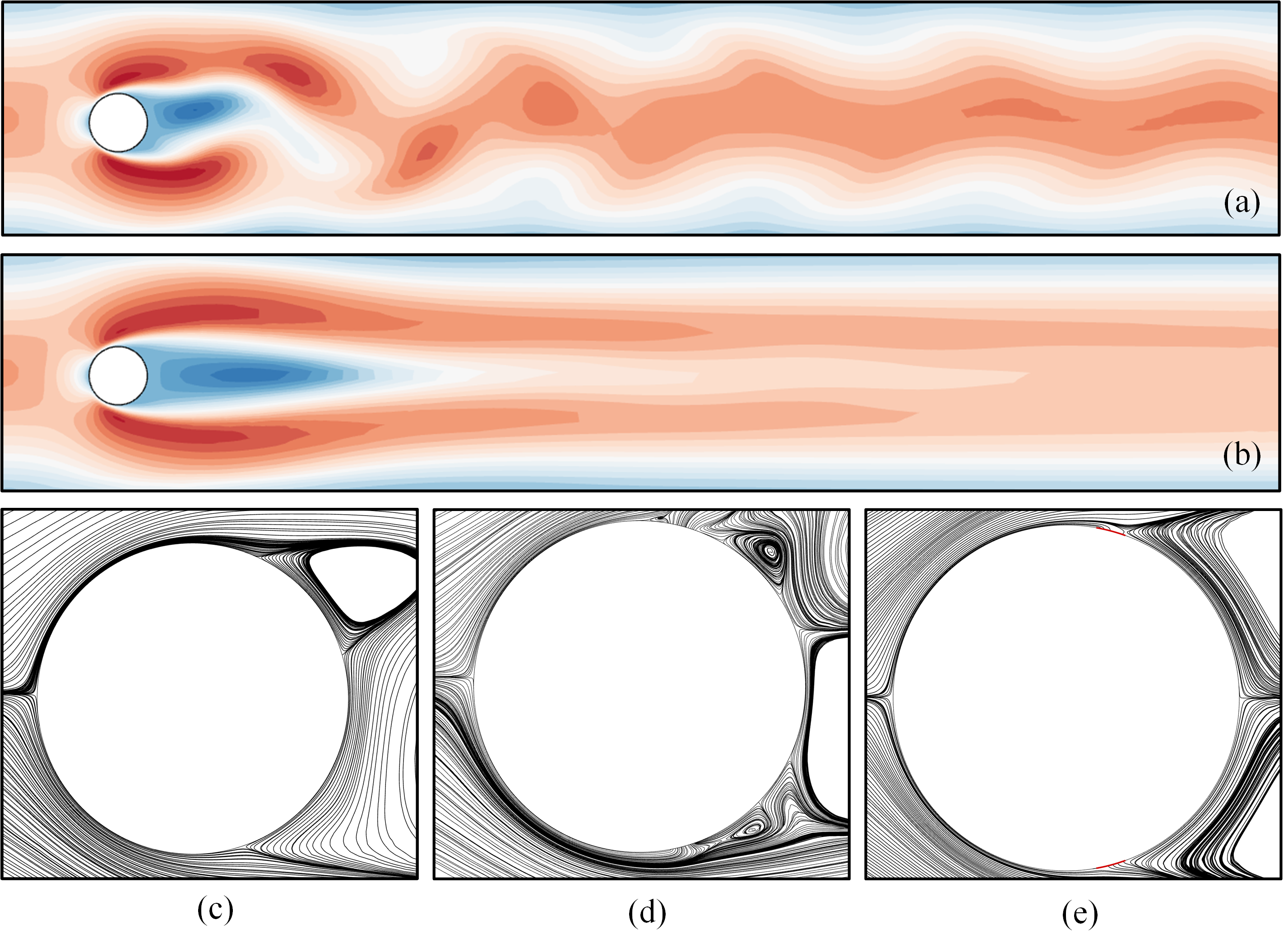}
    \caption{
    (a) Velocity contours of flow around a cylinder in the baseline case. (b) Velocity contours of the controlled flow with synthetic jets positioned at a 105° azimuthal angle.
    (c) Streamlines around the cylinder in the baseline case. (d) Streamlines around the cylinder with active synthetic jets at a 105° azimuthal angle.
    (e) Streamlines around the cylinder after flow control is completed with synthetic jets activated at a 105° azimuthal angle.
    }
    \label{fig:fig19}
\end{figure*}

Figure~\ref{fig:fig19} illustrates the effectiveness of flow control achieved by synthetic jets positioned at 105° ($J_2^1$ and $J_2^2$), presenting instantaneous streamline patterns at three representative time instances: prior to, during, and after the activation of control. In the baseline case shown in figure~\ref{fig:fig19}(a), a classical Kármán vortex street is observed downstream of the cylinder, characteristic of unsteady wake dynamics. Upon activation of the synthetic jets, as depicted in figure~\ref{fig:fig19}(b), the recirculation region on the leeward side of the cylinder expands significantly, and vortex shedding is completely suppressed. Furthermore, figure~\ref{fig:fig19}(c) reveals the baseline flow’s separation zone and the development of a large, closed vortex structure on the rear side of the cylinder, which is effectively mitigated through synthetic jet control.

When the jet positioned at the 105° azimuth is activated, it operates through alternating ejection and suction of the surrounding fluid, effectively altering the local flow without introducing any new mass into the system, as shown in figure \ref{fig:fig19} (d). The synthetic jets located at $J_2^1$ and $J_2^2$ consistently satisfy the condition $Q_1 + Q_2 = 0$, ensuring that the net mass flow in the system remains zero. However, the momentum and hydrodynamic impulse generated by the jets are not zero.
The interaction between the synthetic jets and the crossflow leads to the formation of new vortex pairs. These high-energy vortices convect over the cylinder, locally modifying the streamlines around its surface, as shown in figure \ref{fig:fig19} (e). 
As a result, the apparent aerodynamic shape of the cylinder is significantly improved, along with its aerodynamic characteristics. By altering the local streamlines and inducing noticeable surface modifications, synthetic jets effectively control the flow, reducing drag and improving overall aerodynamic efficiency.
\cite{smith1998} and \cite{amitay} experimentally demonstrated the formation, evolution, and control mechanisms of synthetic jets in flow control. The use of a DRL-based approach reveals a dynamically evolving flow control process that aligns with their findings. This consistency further validates the effectiveness of synthetic jets in altering flow behavior and enhancing aerodynamic performance through active flow control strategies.

In section \ref{sec:sq}, the training results for various jet configurations around the square cylinder demonstrate that positioning synthetic jets near the rear corners of the cylinder yields the optimal control performance, achieving both low energy consumption and high flow control efficiency. In this configuration, drag reduction and lift suppression are significantly enhanced, while vortex shedding in the wake is entirely suppressed with minimal external energy input.
Further analysis of the separation points for both square and circular cylinders, as shown in section \ref{fig:fig18}, combined with the results from section \ref{sec:cy} and section \ref{sec:sq}, indicates that placing synthetic jets near the flow separation points is critical for effective flow control. Only when the jets are positioned in close proximity to these separation points can vortex shedding be fully suppressed. 

\begin{figure*}[htb!]
    \centering
    \includegraphics[width=0.7\textwidth]{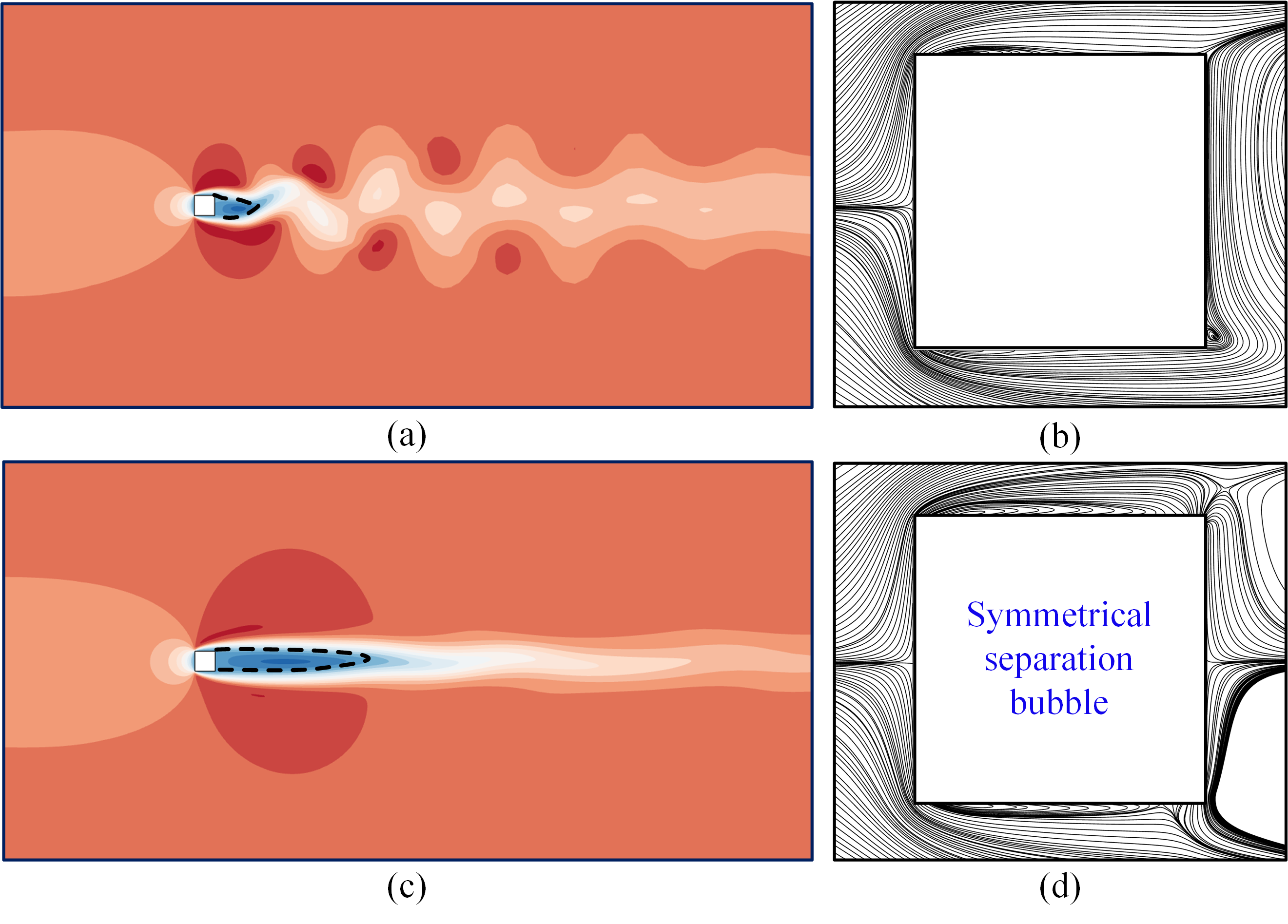}
    \caption{
    (a) Velocity contours of flow around a square cylinder in the baseline case. 
    (b) Streamlines around the square cylinder in the baseline. 
    (c) Velocity contours of the controlled flow with jets positioned at the $J_3$ location.
    (d) Streamlines around the square cylinder after flow control with jets positioned at the $J_3$ location.
    }
    \label{fig:fig20}
\end{figure*}

The comparison between the baseline flow and the controlled flow, with jets positioned at the rear corners ($J_3^1$ and $J_3^2$), is illustrated in figure \ref{fig:fig20}. The placement of the synthetic jets near the rear corners effectively suppresses vortex formation and stabilizes the wake flow. Streamline patterns around the square cylinder reveal significant local alterations near the rear corners due to the action of the synthetic jets. In the controlled flow, the flow separation regions on both sides of the square cylinder become more defined, and two stable, closed separation bubbles form on the cylinder's rear face.
The interaction between synthetic jets and the external flow induces significant changes in local flow patterns and streamlines, facilitating linear momentum transfer without mass addition. The alternating phases of blowing and suction generate vortices that enhance boundary layer mixing, supplying the necessary energy to overcome adverse pressure gradients and enabling complete flow reattachment around the square cylinder.

\section{CONCLUSIONS}\label{sec:Conclusions} 

We use DRL algorithms to train synthetic jet-based flow control strategies for circular and square cylinders. The agent learns to optimize jet actuation to minimize drag and suppress instabilities while balancing energy efficiency and control effectiveness. By systematically evaluating single- and multi-jet configurations, we identify optimal jet placements and provide qualitative guidelines based on hydrodynamic analyses.

\begin{enumerate}
    \item  Flow around a circular cylinder scenario.
    The single-jet training results indicate that positions $ J_1 $ and $ J_5 $ show slower and lower reward function convergence, suggesting lower training effectiveness and energy efficiency. In contrast, positions $ J_2 $, $ J_3 $, and $ J_4 $ achieve rapid convergence and superior control strategy acquisition. Despite all positions achieving an 8\% drag reduction and 99\% lift coefficient suppression, $ J_1 $ and $ J_5 $ require three times the energy of $ J_2 $ to reach the same performance level. 
    For the control strategy involving the simultaneous activation of two pairs of synthetic jets, the configuration utilizing jets at positions $ J_2 $ and $ J_4 $ demonstrates higher energy efficiency compared to the configuration using jets at position $ J_3 $. 
    When three pairs of synthetic jets are activated simultaneously, the control strategy with jets located at $ J_2 $ demonstrates the highest energy efficiency, compared to those at $ J_4 $  or $ J_3 $.

    \item Flow around a square cylinder scenario. 
    When controlling a single pair of synthetic jets, placement at $J_3$ yields the best performance, achieving 14.4\% drag reduction with only 1.7\% of the inlet flow rate—demonstrating both high control effectiveness and energy efficiency. In contrast, jets at $J_1$ require 44\% of the flow rate for just 12.1\% drag reduction, resulting in higher energy cost and lower stability. Using dual jet pairs at $J_2$ and $J_3$ improves force stabilization and achieves performance comparable to $J_3$ alone, but with slower reward convergence. When all three jet pairs $J_2$, $J_3$, $J_4$ are active, $J_3$ still dominates in efficiency and control effectiveness, highlighting its optimal placement.

    \item Mechanism of energy-efficient flow control. 
    DRL training results reveal that placing synthetic jets near 105° on a circular cylinder and at the rear corners of a square cylinder yields optimal energy efficiency and flow control performance. Positioning jets at separation points enhances momentum exchange and induces beneficial vortex structures, promoting boundary layer reattachment without mass addition. This strategic alignment with key flow dynamics significantly improves control effectiveness, stability, and energy efficiency.

\end{enumerate}

The effectiveness of DRL in flow control lies in its ability to autonomously learn optimal strategies through interaction with complex, nonlinear environments. By capturing intricate state–action relationships, DRL enables exploration of control solutions beyond the scope of conventional methods. This study demonstrates that strategic placement of synthetic jets allows for a favorable trade-off between control performance and energy efficiency. Coordinated jet configurations yield synergistic effects, enhancing overall control efficacy. The findings underscore the pivotal role of jet placement in achieving robust, energy-efficient flow control.

\section{APPENDIX I}\label{app1}

To ensure that the numerical results are independent of the mesh resolution, a mesh independence study is conducted for both circular and square bluff bodies. The computational domain is discretized using three levels of mesh density: coarse, medium, and fine. The total number of grid cells for each case is reported in table~\ref{tab:table2}. The drag coefficient $ C_{D} $, lift coefficient $ C_{L} $, and Strouhal number $ St $ are selected as key parameters for evaluation.

\begin{table*}[htb!]
\centering
\caption{Mesh independence test for the flow around a bluff body at $Re=100$.}
\begin{tabular}{
  >{\centering\arraybackslash}p{0.06\textwidth}
  >{\centering\arraybackslash}p{0.2\textwidth}
  >{\centering\arraybackslash}p{0.08\textwidth}
  >{\centering\arraybackslash}p{0.08\textwidth}
  >{\centering\arraybackslash}p{0.1\textwidth}
  >{\centering\arraybackslash}p{0.1\textwidth}
  >{\centering\arraybackslash}p{0.1\textwidth} 
}
\toprule
Bluff & Configuration & Mesh & $C_{D, mean}$ & $C_{D, max}$ & $C_{L, max}$ & $St$ \\
\midrule
         & {Coarse}   & 10,540  & 3.224 & 3.242 & 1.052 & 0.304 \\
cylinder & {Medium}   & 18,484 & 3.205 & 3.225 & 0.990 & 0.300 \\
         & {Fine}     & 26,500 & 3.207 & 3.228 & 0.992 & 0.301 \\
  & \cite{schafer1996benchmark} & - & - & 3.220–3.240 & 0.990–1.010 & 0.295–0.305 \\ 
\midrule
        & {Coarse}   & 8465    & 1.532 & 1.562 & 0.326 & 0.145 \\
square  & {Medium}   & 23 264  & 1.548 & 1.559 & 0.321 & 0.142 \\
        & {Fine}     & 35 179  & 1.549 & 1.561 & 0.320 & 0.141 \\
 & \cite{Singh2009} & - & 1.510 & - & - & 0.147 \\  
 & \cite{Sen2011} & - & 1.529 & - & - & 0.145 \\  
\bottomrule
\end{tabular}
\label{tab:table2}
\end{table*}

For the circular cylinder, the mean drag coefficient, maximum drag coefficient, and maximum lift coefficient all exhibit small variations and converge towards consistent values with increasing mesh resolution. The Strouhal number also remains stable across different mesh resolutions. These results are in close agreement with the reference benchmark study by \citeauthor{Schafer1996}, further validating the numerical setup.
For the square cylinder, a similar trend of convergence is observed for the mean drag coefficient and Strouhal number, with minimal changes across different mesh resolutions. The final values obtained are in excellent agreement with prior studies, such as \citeauthor{Singh2009} and \citeauthor{Sen2011}.
Based on these observations, the medium mesh resolution is deemed sufficient for achieving accurate and computationally efficient results. Therefore, all subsequent simulations are conducted using the medium mesh to balance computational cost and solution accuracy.

\section{APPENDIX II}\label{app2}

\begin{figure*}[htb!]
    \centering
    \includegraphics[width=0.925\textwidth]{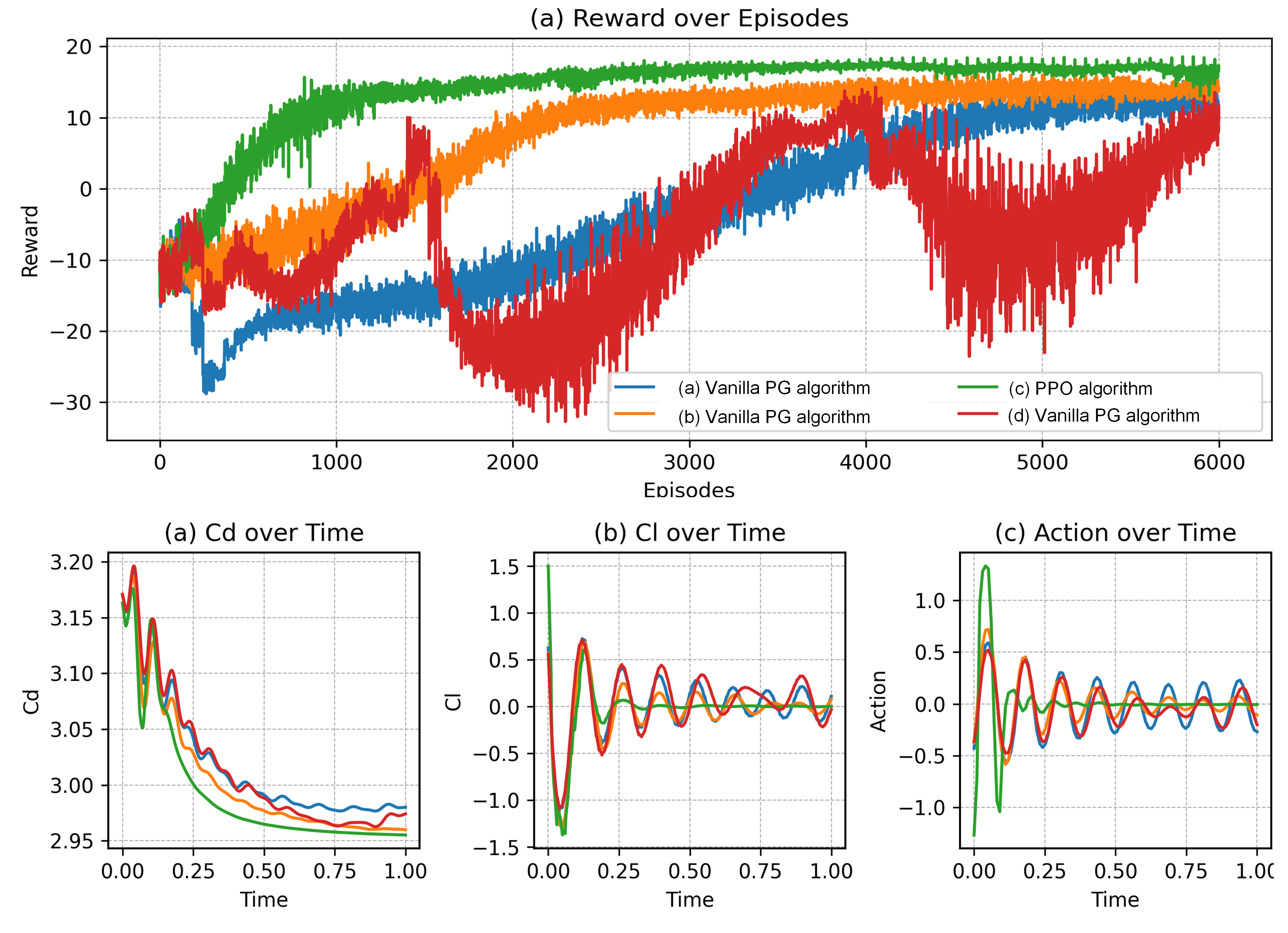}
    \caption{Performance evaluation of different jets configurations for flow control around a cylinder. 
    (a) Reward curves over training episodes for various jets positions. 
    (b) $C_D$ over non-dimensional time, comparing the baseline scenario with different jets configurations. 
    (c) $C_L$ over non-dimensional time, comparing the baseline scenario with different jets.
    }
\label{fig:fig21}
\end{figure*}

The performance of two policy-based reinforcement learning algorithms in active flow control around a circular cylinder is comprehensively compared in figure~\ref{fig:fig21}.
As shown in figure \ref{fig:fig21}(a), the control strategy trained using the PPO algorithm consistently achieves the highest cumulative reward throughout the training process, exhibiting faster convergence and reduced fluctuations. This indicates superior learning efficiency and control performance. In contrast, the three variants of the vanilla policy gradient algorithm (curves a, b, and d) converge more slowly and exhibit significantly higher variance, reflecting unstable learning dynamics and suboptimal control effectiveness.

The time-resolved aerodynamic performance of the learned controllers is further assessed in terms of the drag and lift coefficients. Figure \ref{fig:fig21}(b) shows the evolution of the drag coefficient $C_D$, where all control strategies yield a gradual decrease in $C_D$ toward a lower steady-state value compared to the uncontrolled baseline. The PPO-based controller achieves the most rapid and pronounced drag reduction, with $C_D$ dropping below 2.98. This result highlights the efficacy of PPO in mitigating wake dynamics and suppressing vortex-induced drag.
Figure \ref{fig:fig21}(c) presents the temporal evolution of the lift coefficient $C_L$. The PPO-based strategy exhibits the fastest decay of $C_L$ oscillations, achieving near-zero mean fluctuations within a short time window, thereby indicating improved wake stabilization. In contrast, the Vanilla PG-based controllers show slower decay rates and larger residual oscillations, suggesting inferior suppression of unsteady lift forces.

In summary, the PPO algorithm demonstrates clear advantages over the Vanilla PG approaches in both training stability and flow control performance. It facilitates faster policy acquisition, more effective drag suppression, and enhanced lift stabilization. The consistent improvements observed in reward convergence, drag reduction, and lift damping collectively confirm the effectiveness and robustness of PPO-based strategies for active flow control applications.

\section{APPENDIX III}\label{app3}

To enhance the reproducibility of the present study, the principal parameters employed in both the CFD simulations and the DRL training procedure are summarized in table~\ref{tab:table3}. The simulations are conducted at a Reynolds number of $Re = 100$, with a non-dimensional time step of $5 \times 10^{-4}$. To ensure efficient data generation and throughput, 60 parallel CFD environments are instantiated, each bound to a single CPU core, leading to a total allocation of 61 CPU cores. The deep reinforcement learning agent, based on the PPO algorithm, is trained over 3000 episodes. The relevant parameters of the PPO algorithm are summarized in the table.

\begin{table*}[hbt!]
\centering
\caption{Parameters of both the CFD simulation and the DRL algorithm.}
\begin{tabular*}{\textwidth}{@{\extracolsep{\fill}} l l l }
\toprule
Parameter & Symbol & Value \\
\midrule
Reynolds number & $Re$ & 100  \\  
Numerical time step (non-dimensional) & dt & $5 \times 10^{-4}$  \\ 
Parallel CFD environments & - & 60  \\ 
CPUs/environment & - & 1  \\ 
Total CPUs  & - & 61  \\
\midrule
Episode number & - & 3000  \\
Learning rate & lr & 0.001  \\ 
Discount factor & $\gamma$ & 0.97  \\ 
Policy network & $\pi_\theta$ & 512*512  \\ 
Policy Ratio Clipping & $\epsilon$ & 0.3  \\ 
Optimizer  & - & Adam \\ 
Batch size & - & 64  \\
\bottomrule
\end{tabular*}
\label{tab:table3}
\end{table*}


\bibliographystyle{elsarticle-num-names} 
\bibliography{elsarticle}

\end{document}